\documentclass[%
 aps,
 prx,
 reprint,
 superscriptaddress,
 amsmath,amssymb,
 floatfix,
]{revtex4-2}

\usepackage{graphicx}
\usepackage{dcolumn}
\usepackage{multirow,tabularx}%
\usepackage{makecell}
\usepackage{amsfonts}%
\usepackage{mathrsfs}%
\usepackage{xcolor}%
\usepackage{textcomp}%
\usepackage{bm}
\usepackage{booktabs}%
\usepackage{siunitx}
\usepackage{braket}
\usepackage[english]{babel}
\usepackage[hidelinks]{hyperref}

\renewcommand{\selectlanguage}[1]{}
\DeclareSIUnit\dBm{dBm}
\DeclareSIUnit{\rad}{rad}
\DeclareSIUnit{\sample}{S}

\begin{document}

\title{Cavity-assisted dynamic nuclear polarisation}

\author{Tristan Lorriaux}
\thanks{These authors contributed equally to this work.}
\affiliation{ENS de Lyon, CNRS, LPENSL, UMR5672, 69342 Lyon Cedex 07, France}

\author{Yutian Wen}
\thanks{These authors contributed equally to this work.}
\affiliation{ENS de Lyon, CNRS, LPENSL, UMR5672, 69342 Lyon Cedex 07, France}
\affiliation{Quantronics group, Universit\'e Paris-Saclay, CEA, CNRS, SPEC, 91191 Gif-sur-Yvette Cedex, France}
\altaffiliation[Present address: ]{Department of Physics and Astronomy, University of Notre Dame, Notre Dame, Indiana 46556, USA}

\author{Patrice Bertet}
\affiliation{Quantronics group, Universit\'e Paris-Saclay, CEA, CNRS, SPEC, 91191 Gif-sur-Yvette Cedex, France}

\author{Benjamin Huard}
\affiliation{ENS de Lyon, CNRS, LPENSL, UMR5672, 69342 Lyon Cedex 07, France}

\author{Audrey Bienfait}
\email{audrey.bienfait@ens-lyon.fr}
\affiliation{ENS de Lyon, CNRS, LPENSL, UMR5672, 69342 Lyon Cedex 07, France}

\begin{abstract}
Dynamic nuclear polarisation transfers polarisation from electron to nuclear spins by relying on a dissipative channel. Here we introduce a cavity-based DNP mechanism which uses the radiative relaxation of the electronic spin through the Purcell effect to bring about the nuclear spin polarisation. We implement this protocol on an ensemble of bismuth donor spins (I=9/2, S=1/2) in isotopically purified silicon coupled to a superconducting resonator. A single cycle, combining adiabatic microwave and radio-frequency pulses with Purcell relaxation, increases a target ground state's population by a factor of $\qty{1.79 \pm 0.04}{}$; a cumulative enhancement of $\qty{4.6\pm0.2}{}$ is reached with additional cycles, to be compared with the theoretical limit of ninefold set by the degeneracy of the ground manifold. We explain this difference by the competition between Purcell and intrinsic relaxation on the one hand and a collective effect that slows down the electronic spin radiative decay as the nuclear spin population grows on the other hand. Our results place radiative DNP as an available mechanism at millikelvin temperatures where phonon-mediated relaxation is frozen.
\end{abstract}

\maketitle
\section{Introduction}\label{sec:intro}

The polarisation of a nuclear spin ensemble governs the effective number of spins contributing to an experiment. Yet under typical experimental conditions, their transition energies are far below the thermal energy, so that even at low cryogenic temperatures, the polarisation of a nuclear spin transition remains small~\cite{abragam1986,slichter1990}.
This low polarisation sets the detection sensitivity of nuclear magnetic resonance (NMR)~\cite{hoult1976}. It also limits spin-ensemble quantum memories, in which electron--nuclear spins are coupled to a cavity: an unpolarised ensemble spreads its population over many states, so that only a fraction of the spins interact resonantly with the microwave field~\cite{kubo2010,schuster2010}, while the remaining spins form a fluctuating bath that shortens the coherence time~\cite{klauder1962,takahashi2008,diniz2011}.

Nuclear polarisation beyond thermal equilibrium is most usually achieved by accessing a higher-polarisation reservoir, provided a suitable dissipative thermalisation mechanism exists. This is the case when the spin ensemble is pre-polarised at lower temperatures, or equivalently at higher magnetic field~\cite{hirsch2015}. This cooling can also be engineered via spin-selective optical transitions~\cite{walker1997,lampel1968, yang2009}, or by exploiting nuclear spin exchange interaction~\cite{meier2018,pound1950} as in parahydrogen-induced polarisation~\cite{bowers1986,bowers1987,adams2009}. Polarisation can also be achieved by measurement, by relying on a projective readout followed by post-selection or by a conditional correction, as demonstrated on individual nuclear spins coupled to an electronic spin~\cite{neumann2010,robledo2011a,pla2013,osullivan2025} or by quantum feedback techniques~\cite{wang2001}. Dynamic nuclear polarisation (DNP) techniques couple nuclear spins to electronic spins to exploit the larger polarisation of the latter~\cite{abragam1978,dai2025} and cool down the nuclear spins in a timescale set by the dynamics of the electronic spin relaxation. In DNP experiments aimed at enhancing NMR sensitivity, this motivates efforts to identify and engineer electron–nuclear systems with favourable relaxation properties~\cite{can2015,corzilius2020,hu2004,sauvee2013}, to introduce electron spins chemically~\cite{ward1972,dai2025}, and to prepare the electron spin ensemble optically~\cite{henstra1990,liu2017} or through temperature and field cycling, as in dissolution DNP~\cite{ardenkjaer-larsen2003}. 

In the context of microwave-based quantum memories, where electro-nuclear spin systems are coupled to quantum circuits, in experiments realized at millikelvin temperature, dissipation makes these DNP protocols challenging since phonon-mediated relaxation freezes out. Here, we develop a DNP scheme based on the coupling to a cold electromagnetic bath. When a spin transition is brought into resonance with a microwave cavity, the Purcell effect~\cite{purcell1946} can enhance its spontaneous emission rate by orders of magnitude, making radiative decay the dominant relaxation channel. This approach has previously been applied to efficiently reset electronic spins~\cite{bienfait2016}, to achieve electronic hyperpolarisation~\cite{albanese2020}, and has also enabled single-spin detection~\cite{wang2023} through the use of microwave photon detectors~\cite{lescanne2020}. 

We use this frequency-selective decay to engineer a DNP mechanism for bismuth donors in isotopically purified silicon coupled to a superconducting resonator. This species possesses 9 ground states that are weakly polarised at the millikelvin temperature at which we operate. Through a combination of drives and Purcell relaxation, we successively transfer the populations of every state to a single one, yielding a population enhancement of $\qty{4.6\pm0.2}{}$ for the target state compared to its initial value - approaching the theoretical limit of 9. We identify and quantitatively explain the mechanisms that limit hyper-polarisation: competition with intrinsic relaxation, imperfect inversion, and a slow-down of the radiative relaxation due to cavity-induced spin interactions as the nuclear spin population grows and the system comes close to the strong spin-cavity interaction regime. Our results establish Purcell engineering as a route to dynamic nuclear polarisation. In the case of a quantum memory based on electro-nuclear spin system, it enables designs of higher cooperativity or operation with more dilute spin ensembles, which is beneficial in the quest for longer coherence times. In the future, applying this technique to a bath of nearby nuclear spins with weaker coupling to the electronic spin would suppress spectral diffusion and lengthen coherence times, potentially without requiring an isotopically purified host.

\begin{figure}[htpb!]
    \centering
    \includegraphics[width=\columnwidth]{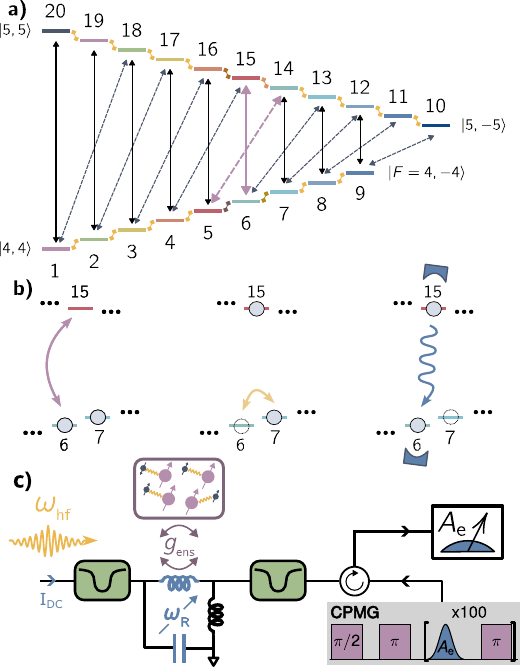}
    \caption{\label{fig:fig1}\textbf{Purcell-DNP principle and experimental setup} \textbf{a)} Energy level diagram of Bi:Si donors at low magnetic field (not to scale, see main text). The two transitions addressable by the resonator at $\qty{7.4}{\giga\hertz}$ are in purple. Hyperfine transitions (yellow) near $\qty{0.2}{\giga\hertz}$ connect neighbouring energy levels inside the ground and excited manifolds. \textbf{b)} Single cycle of the Purcell-DNP protocol. An electronic-like swap between $|6\rangle$ and $|15\rangle$ is followed by a hyperfine swap between $|7\rangle$ and $|6\rangle$. The resonator creates a selective radiative decay channel from $|15\rangle$ to $|6\rangle$ completing the pumping cycle. \textbf{c)} Experimental circuit: A frequency-tunable superconducting resonator (blue) is coupled with strength $g_\mathrm{ens}$  to a Bi:$^{28}$Si spin ensemble.     
      Notch-filters (green) ensure the resonator has a high-quality factor enabling the Purcell effect, while leaving the sub-gigahertz (hf) transitions addressable and the resonator tunable in frequency through the application of a DC current $I_{\mathrm{DC}}$ modulating its kinetic inductance (see Appendix~\ref{m:device}~\cite{wen2025}). Gigahertz transitions are probed through the resonator using CPMG sequences.}
\end{figure}

\section{Protocol}

 Bismuth spins have an electronic spin $S=1/2$ and nuclear spin $I=9/2$, and their energy structure is governed by Zeeman interactions for both the electronic and nuclear spins (gyromagnetic ratios $\gamma_\mathrm{e}= \qty{27.997}{\giga\hertz\per\tesla}$ and $\gamma_n = \qty{6.962}{\mega\hertz\per\tesla}$)  and an isotropic hyperfine interaction of strength  $A=\qty{1.475}{\giga\hertz}$. In a static magnetic field $\mathbf{B}_0$, it is described by the Hamiltonian $\hat{H}/h = \gamma_{\mathrm{e}}\, \mathbf{B}_0\cdot\hat{\mathbf{S}} - \gamma_{\mathrm{n}}\, \mathbf{B}_0\cdot\hat{\mathbf{I}} + A\, \hat{\mathbf{S}}\cdot\hat{\mathbf{I}}$.
In the low field regime $\gamma_e B_0 \ll A$, the 20 energy eigenstates are hybridized electron–nuclear spin states labelled by the total angular momentum quantum numbers $F=S+I$ and $m_{\mathrm{F}}$.
They split into a ground manifold ($F=4$, nine states) and an excited manifold ($F=5$, eleven states), see Fig.~\ref{fig:fig1}a. We identify eigenstates by ascending energy as $|i\rangle$. Between the two manifolds, 18 electronic-like transitions with $\Delta m_{\mathrm{F}}=\pm1$ occur at gigahertz frequencies. Within each manifold, hyperfine transitions connect neighbouring states with frequencies near $\qty{200}{\mega\hertz}$ at the field $B_0=\qty{65.37}{\milli\tesla}$ we operate at. At the thermal equilibrium corresponding to our cryostat temperature $T=\qty{25}{\milli\kelvin}$, only the ground manifold is occupied and each nuclear transition has a weak polarisation $p_{\mathrm{th}}\approx0.19$. 
At the beginning of each experimental sequence, we apply a scrambling sequence enabling to average out the population within the ground manifold to $p_\mathrm{avg}=1/9$ (see Appendix~\ref{m:reset}) effectively initializing the polarisation of each nuclear transition to $0$.

The Purcell-DNP protocol exploits the radiative relaxation enabled by a narrow linewidth cavity to reset the electronic spins and operates in three steps, as illustrated in Fig.~\ref{fig:fig1}b. First, a microwave adiabatic pulse transfers population from ground state $|6\rangle$ to the excited state $|15\rangle$. Second, an hf pulse swaps the population of a neighbouring ground state (e.g. $|7\rangle$) into the now-empty state $|6\rangle$. Third, the resonator is tuned into resonance with $|6\rangle\leftrightarrow|15\rangle$, enabling the population in $|15\rangle$ to relax radiatively back to $|6\rangle$ at the Purcell rate $\Gamma_\mathrm{P}$. In the absence of other intrinsic decay mechanisms, the final population in $|6\rangle$ doubles. Additional cycles containing more nuclear swap pulses can be used to hyperpolarise the full ground manifold and reach a factor 9 enhancement.

\section{Device}

\begin{figure}[htpb!]
    \centering
    \includegraphics[width=\columnwidth]{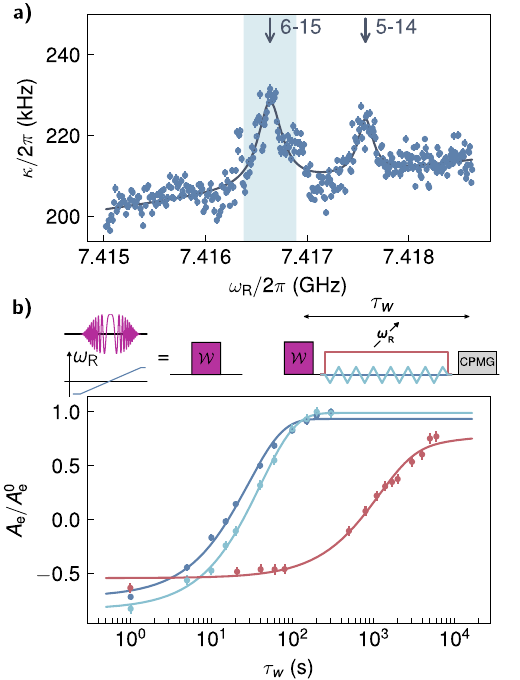}
    \caption{\label{fig:fig2} \textbf{Spin ensemble characterization.} \textbf{a)} Resonator decay rate $\kappa$ as a function of its frequency. When tuned through the $|6\rangle\leftrightarrow|15\rangle$ and $|5\rangle\leftrightarrow|14\rangle$ transitions, the spin ensemble induces additional losses. Lorentzian fits yields the transition frequencies, linewidths, and cooperativities (see main text). 
    \textbf{(b)} Inversion-recovery measurement of the $\ket{6}\leftrightarrow\ket{15}$ lifetime, using an adiabatic WURST inversion pulse. With the resonator on resonance, Purcell relaxation dominates and sets $T_1 = \qty{32(2)}{\second}$ (blue). Sweeping the resonator frequency by $\qty{\pm 150}{\kilo\hertz}$ during the relaxation step broadens the Purcell reset bandwidth, at the price of extending the effective lifetime to $T_1^{\mathrm{eff}} = \qty{41(1)}{\second}$ (cyan). With the resonator detuned, intrinsic relaxation dominates (red); the solid line is a fit accounting for the two decay paths of $\ket{15}$, $\Delta m_F = \pm 1$ towards $\ket{6}$ and $\ket{4}$ (see Appendix~\ref{m:intrinsicT1}). Circles are data, solid lines are fits; the two Purcell curves are extrapolated beyond the measured range.}
\end{figure}

In our experiment, the cavity is implemented by a planar microwave superconducting resonator whose design and characterization are detailed in~\cite{wen2025}. At the magnetic field we operate, the superconducting resonator can be tuned through the $|6\rangle\leftrightarrow|15\rangle$ and $|5\rangle\leftrightarrow|14\rangle$ doublet by tuning its kinetic inductance via a DC current~\cite{asfaw2017,wen2025} from $\omega_\mathrm{R}/(2\pi)=\qty{7.420}{\giga\hertz}$ to $\qty{7.349}{\giga\hertz}$. The Bismuth atoms are implanted on a $\qty{600}{\nano\meter}$-deep layer at the surface of a $^{28}$Si chip placed $\qty{1.1}{\micro\meter}$ above the resonator inductive wire, see Appendix~\ref{m:device}. Each spin couples with strength $g$ to the resonator through a Jaynes-Cummings interaction, cumulating in an ensemble coupling $g_{\mathrm{ens}}^2 = \int \rho(g) g^2 dg$ where $\rho(g)$ is the distribution of single-spin couplings (see Appendix~\ref{m:CPMG}). When the resonator is tuned through either transition (Fig.~\ref{fig:fig2}a), its decay rate increases beyond its bare value $\kappa/2\pi \sim \qty{210}{\kilo \hertz}$ owing to the resonant absorption of microwave photons by the spins at an extra rate $\kappa_\mathrm{s} =4 g_{\mathrm{ens}}^2/\Gamma$, with $\Gamma$ the spin full-width half-maximum linewidth. Fitting the absorption peaks to Lorentzians yields $\omega_{6-15}/(2\pi)=\qty{7.416635}{\giga\hertz}$, respective linewidths $\Gamma_{6-15}/(2\pi)=\qty{288\pm24}{\kilo\hertz}$ and $\Gamma_{5-14}/(2\pi)=\qty{182\pm27}{\kilo\hertz}$, and a frequency splitting of $\qty{940\pm11}{\kilo\hertz}$ between the two transitions. The corresponding ensemble cooperativities $C=\kappa_\mathrm{s} /\kappa$~\cite{staudt2012} are $C^{6-15} =\qty{0.1 \pm 0.01}{}$ and $C^{5-14} = \qty{0.06\pm 0.01}{}$, reflecting the different spin matrix elements of the two transitions (see Appendix~\ref{m:hamiltonian}), and are close to the state of the art for this spin species~\cite{osullivan2020}. Importantly, the spin linewidth $\Gamma$ is smaller than the doublet splitting, ensuring that each transition can be addressed independently. Such narrow linewidth is realized in our experiment by using a flip-chip technique that leaves the spin chip free of metal-induced strain-broadening~\cite{pla2018strain}.

Both cooperativities place the system in the low cooperativity regime, where the Purcell effect can occur and the spin ensemble can be probed by pulsed electron spin resonance. Since the spin ensemble has a long coherence time ($T_2=\qty{3.4}{\milli\second}$), we use Carr-Purcell-Meiboom-Gill (CPMG) sequences to measure the polarisation of the $|6\rangle\leftrightarrow|15\rangle$ transition in a single-shot measurement. An initial $\pi/2$ pulse followed $\qty{250}{\micro\second}$ later by a train of $100$ $\pi$ pulses gives rise to $100$ echoes that we measure through heterodyne detection (see Appendix~\ref{m:setup}). Averaging the echoes together yields an average signed echo amplitude $A_\mathrm{e}$ that is proportional to the difference in population between $|15\rangle$ and $|6\rangle$ for spins lying within a $\qty{40}{\kilo\hertz}$ window around the drive frequency $\omega_\mathrm{d}$ (see Appendix~\ref{m:CPMG}). The drive amplitudes of the $\pi/2$ and $\pi$ pulses are chosen to maximize the echo signal. We define $A_\mathrm{e}^0$ as the echo we obtain without any population transfer after a Purcell reset step emptying $|15\rangle$, see Appendix~\ref{m:reset}; any increase in population in $|6\rangle$ can thus be tracked by measuring $A_\mathrm{e}/A_\mathrm{e}^0>1$.

We verify Purcell-dominated relaxation via inversion recovery on the $|6\rangle\leftrightarrow|15\rangle$ transition, see Fig.~\ref{fig:fig2}b. With the resonator on resonance, we measure an exponential decay of characteristic time $T_1=\qty{32 \pm 2}{\second}$; detuning the resonator reveals that state $|15\rangle$ has an intrinsic lifetime of $T_1^{\mathrm{int}}=\qty{24\pm2}{\minute}$ (see Appendix~\ref{m:intrinsicT1}), similar to \cite{bienfait2016} and confirming that Purcell decay dominates by nearly two orders of magnitude. From the on-resonance $T_1$ and the single-spin Purcell rate formula $\Gamma_{\mathrm{P}}= \kappa g^2/[\kappa^2/4+\delta^2]$, we extract a mean coupling $\bar{g}/(2\pi)=\qty{16.3 \pm 0.5}{\hertz}$ for the spins contributing to the echo. Since the spin linewidth exceeds the resonator linewidth ($\Gamma/\kappa\sim 1.5$), spins away from resonance are less efficiently reset by the Purcell effect. To mitigate this effect, we modulate the resonator frequency by $\pm\qty{150}{\kilo\hertz}$ during the relaxation step (see Appendix~\ref{m:wigglyPurcell}) broadening the effective reset bandwidth at the cost of a longer effective $T_1^{\mathrm{eff}}=\qty{41 \pm 1}{\second}$.

\section{Population transfer}\label{sec:wurst}
To transfer population between states, we use WURST (wideband, uniform rate, smooth truncation) adiabatic inversion pulses~\cite{odell2013,osullivan2022} which are resilient to both the coupling inhomogeneity of our micro-resonator geometry and the spin frequency inhomogeneity (see Appendix~\ref{m:WURST}). To make the application of a chirped pulse compatible with the narrow linewidth of the resonator, we dynamically tune the resonator frequency to the instantaneous drive frequency during the pulse. For hf transitions, no such correction is needed as the resonator couples to the drive line by a stop-band filter with unit-transmission below $\qty{4}{\giga\hertz}$~\cite{wen2025}.

\begin{figure}[htpb!]
    \centering
\includegraphics[width=\columnwidth]{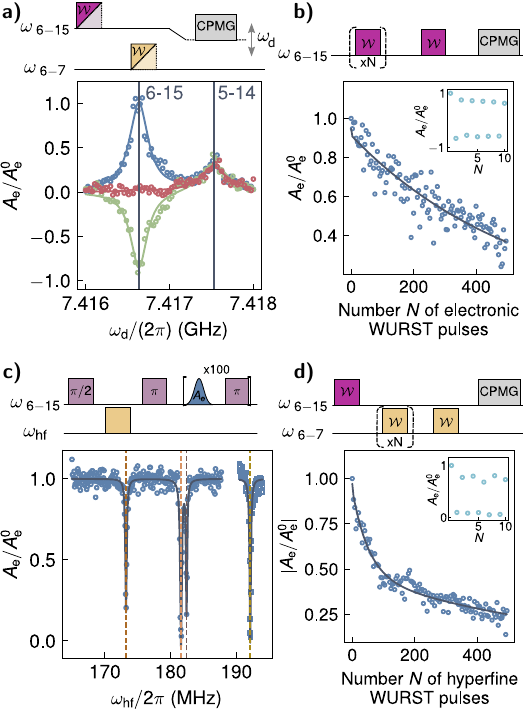}
\caption{\label{fig:fig3} \textbf{Population transfer characterization.} 
\textbf{a)} CPMG spectroscopy of $|6\rangle\leftrightarrow|15\rangle$ and $|5\rangle\leftrightarrow|14\rangle$ (blue). Applying a WURST pulse resonant with $|6\rangle\leftrightarrow|15\rangle$
inverts that transition while leaving $|5\rangle\leftrightarrow|14\rangle$ unchanged  (green), confirming transition selectivity. Applying a nuclear WURST pulse on $|6\rangle\leftrightarrow|7\rangle$ afterwards equilibrates the $|6\rangle\leftrightarrow|15\rangle$ population to zero net polarisation (red), while $|5\rangle\leftrightarrow|14\rangle$ remains unaffected. Solid lines are Lorentzian fits.
\textbf{b)} Normalized echo amplitude at $\omega_{\mathrm{d}} = \omega_{6\text{-}15}$ after $N$ electronic WURST pulses, $N$ a multiple of four (circles). A fit to the data (see Appendix~\ref{m:WURST})  yields an inversion efficiency $\eta_e=0.9996\pm0.0005$. Inset: consecutive values of $N$.
\textbf{c)} ENDOR spectroscopy: a nuclear pulse at frequency
$\omega_{\mathrm{hf}}$ inserted into the CPMG sequence reveals four nuclear
transitions (dashed lines, colour-coded as in Fig.~\ref{fig:fig1}b), from which
we determine the central frequency of the nuclear WURST pulses.
\textbf{d)} Normalized echo amplitude after inverting  $|6\rangle\leftrightarrow|15\rangle$ and applying an even number $N$ of WURST pulses on $|6\rangle\leftrightarrow|7\rangle$ (circles). A fit to the data (see Appendix~\ref{m:WURST}) yields the nuclear inversion efficiency $\eta_n=0.991\pm0.001$ (solid line). Inset: consecutive values of $N$.}
\end{figure}

The chirp range $\delta_W=\pm\qty{250}{\kilo\hertz}$ of our pulses is chosen to invert the targeted
microwave transition $\ket{6}\leftrightarrow\ket{15}$ without addressing the other transition of the doublet $\ket{5}\leftrightarrow\ket{14}$ (see Fig.~\ref{fig:fig3}a). We use the areas of the peaks measured with CPMG spectroscopy with and without the inverting pulse to determine that $\qty{
90 \pm 3}{\percent}$ of the spins are inverted. The change in area of $\ket{5}\leftrightarrow\ket{14}$ is at most \qty{11}{\percent}, consistent with it being a spectator transition. Concentrating on the spins lying at the center of the $\ket{6}\leftrightarrow\ket{15}$ transition, we estimate the inversion efficiency of the electronic pulse by measuring the spin polarisation at $\omega_{\mathrm{d}} = \omega_{6\text{-}15}$ after $N$ chirped
pulses (Fig.~\ref{fig:fig3}b). The polarisation alternates in sign, and decays as
$P_{6\text{-}15} = A_{\mathrm{e}}/A_{\mathrm{e}}^{0} = (1-2\eta_{\mathrm{e}})^{N}$, yielding an near-perfect inversion efficiency $\eta_{\mathrm{e}} = 0.9996\pm0.0001$ for a single WURST pulse. The difference with the inversion efficiency determined by integrating the entire line indicates that the spins far detuned are not as well inverted  (see Appendix~\ref{m:fidelities}).

For nuclear transitions, we first identify the relevant transition frequencies via ENDOR spectroscopy~\cite{schweiger2001} (see Fig.~\ref{fig:fig3}c) then characterize nuclear population transfer on the $|6\rangle\leftrightarrow|7\rangle$ transition using 
$\delta_W = \pm\qty{150}{\kilo\hertz}$. We begin by performing an inversion on $|6\rangle\leftrightarrow|15\rangle$ before swapping $|6\rangle\leftrightarrow|7\rangle$, thus equalizing the ground and excited populations on the $|6\rangle\leftrightarrow|15\rangle$ transition. Performing a CPMG spectroscopy of the doublet (see red curve in Fig.~\ref{fig:fig3}a), we indeed observe no remaining polarisation on the $|6\rangle\leftrightarrow|15\rangle$ transition while the $|5\rangle\leftrightarrow|14\rangle$ transition is left unaffected, indicating that the nuclear population is entirely transferred. By increasing the number $N$ of WURST pulses $\mathcal{W}_{6-7}$ applied on $|6\rangle\leftrightarrow|7\rangle$ transition, we expect to measure an echo amplitude evolving as $P_{6-15}=A_\mathrm{e}/A_\mathrm{e}^0 =-((1-2\eta_n)^N+1)/2$ where $\eta_n$ is the nuclear inversion fidelity. As shown in Fig.~\ref{fig:fig3}d, the polarisation stays constant over tens of
pulses before decaying, and converges to $\qty{0.24}{}$ rather than $\qty{0.5}{}$. We
attribute this floor to a slow leakage of population out of $\ket{6}$ and
$\ket{7}$ into the other ground states, which accumulates over experimental
repetitions (see Appendix~\ref{m:fidelities}). Fitting the full level-population model, with the intrinsic decay rates fixed to their
independently measured values (see Appendix~\ref{m:intrinsicT1}), gives a nuclear inversion probability $\eta_{\mathrm{n}} = \qty{0.991 \pm 0.001}{}$.

\section{Nuclear pumping}\label{sec:dnp1}

\begin{figure}[htpb!]
    \centering
    \includegraphics[width=\columnwidth]{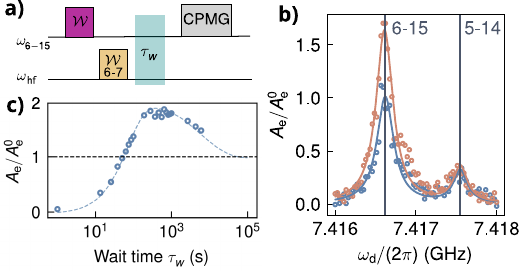}
    \caption{\label{fig:fig4}  \textbf{Observation of radiative nuclear pumping.}
    \textbf{a)} DNP pulse sequence for $1$ cycle: one electronic and one nuclear WURST pulses followed by a Purcell recovery wait time $\tau_W$ realizes the scheme described in Fig.~\ref{fig:fig1}b. \textbf{b)} CPMG spectroscopy of the $|6\rangle\leftrightarrow|15\rangle$ transition before (blue) and after (orange) one DNP cycle with $\tau_W = \qty{300}{s}$. Solid lines are Lorentzian fits. \textbf{c)} Normalized echo amplitude at $\omega_\mathrm{d} = \omega_{6-15}$ as a function of $\tau_W$. After an optimal recovery time $\tau_W = \qty{280}{\second}$, the polarisation enhancement reaches $\qty{1.79\pm0.04}{}$ before decaying on a timescale set by hyperfine diffusion. Dashed line is a rate equation fit, see Appendix~\ref{m:DNPODE}.}
\end{figure}

We first demonstrate a single cycle of the Purcell-DNP protocol. Using inversion pulses on $|6\rangle\leftrightarrow|15\rangle$ and $|6\rangle\leftrightarrow|7\rangle$ followed by Purcell-enhanced relaxation during a delay $\tau_W$ transfers some of the spins initially in $|7\rangle$ into $|6\rangle$  (Fig.~\ref{fig:fig4}a). After an optimal delay $\tau_W = \qty{280}{\second}$, the echo measured at $\omega_\mathrm{d} = \omega_{6-15}$  increases by a factor of $\qty{1.79\pm0.04}{}$
(Fig.~\ref{fig:fig4}c), confirming an increased number of spins in $|6\rangle$. The echo then remains at this level for about $\qty{700}{\second}$ before decaying back to its unpumped level. We attribute this decay and the gap between the observed factor of $\qty{1.79 \pm 0.04}{}$ and the expected factor 2 to hyperfine diffusion which creates population leakage from $|6\rangle$ into neighbouring states $|7\rangle$ and $|5\rangle$ (see Appendix~\ref{m:intrinsicT1}). 
A CPMG spectroscopy performed after one cycle (Fig.~\ref{fig:fig4}b) shows a similar increase in the $|6\rangle\leftrightarrow|15\rangle$ peak area of a factor $\qty{1.7 \pm 0.2}{}$ with no change to the $|5\rangle\leftrightarrow|14\rangle$ peak. 

\begin{figure}[htpb!]
    \centering
    \includegraphics[width=\columnwidth]{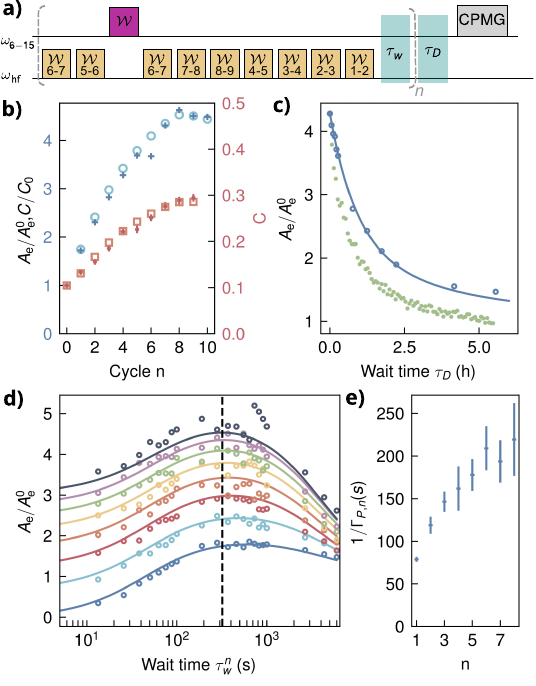}
    \caption{\label{fig:fig5}  \textbf{Complete nuclear pumping and limitations.} 
    \textbf{a)} Pulse sequence for pumping the entire ground manifold population into state $|6\rangle$: a train of WURST pulses brings iteratively the population of one ground state after another to $|15\rangle$ from which it relaxes to $|6\rangle$ after a wait time $\tau_W$, see Appendix~\ref{m:DNPscheme}.
    \textbf{b)} Efficiency after $n$ DNP cycles, by CPMG spectroscopy (blue crosses), or by cw spectroscopy (red points, right axis is in absolute coordinates) after a wait time $\tau_W^n=\qty{320}{\second}$. Circles and squares are from a numerical model capturing the Purcell breakdown effect,  see Appendix~\ref{m:DNPnumerics}.  
    \textbf{c)} Increasing the wait time $\tau_D$ after a 8-pumping scheme leads to the accumulated population in $|6\rangle$ after 8 DNP cycles relaxing (blue points) towards its equilibrium value through hyperfine diffusion (solid line, see Appendix~\ref{m:intrinsicT1}). Repeating CPMG measurements every 4~min (green points) after the DNP peak accelerates this decay.
    \textbf{d)} Normalized echo amplitude at $\omega_\mathrm{d} = \omega_{6-15}$ as a function of the Purcell recovery time $\tau_W^n$, after having completed $n-1$ cycles with wait time $\tau_W=\qty{320}{\second}$ (dashed line). Solid line is the same numerical model (see Appendix~\ref{m:DNPnumerics}). Each color corresponds to a given $n$ from 1 to 8.
    \textbf{e)} Effective Purcell rate determined by fitting each recovery curve separately, showing a slow down of the Purcell effect.}
\end{figure}

Hyperpolarisation across the entire ground manifold can be realized by running iteratively $n = 1, \ldots, 8$ cycles of the DNP sequence shown in Fig.~\ref{fig:fig5}a, using successive hf pulses to shuttle population from all ground states stepwise into $|6\rangle$ at each cycle (see Appendix~\ref{m:DNPscheme}). This specific scheme is chosen to minimize the number of spins transferred to state $|15\rangle$, reducing the protocol sensitivity to the competition between the Purcell effect and other electronic relaxation processes. The cumulative echo enhancement $A_\mathrm{e}/A_\mathrm{e}^0$ on transition $|6\rangle\leftrightarrow|15\rangle$ after each cycle is shown in Fig.~\ref{fig:fig5}b, reaching $\qty{4.6\pm0.2}{}$ after eight cycles (for a total duration of $\qty{45}{\minute}$) before saturating for additional pumping cycles, indicating a significantly larger population in $|6\rangle$. 
DNP is often characterized by an enhancement parameter $E$ corresponding to the increase in signal-to-noise when probing the nuclear transition~\cite{hausser1968,abragam1978}. In our case, $E$ can be estimated by comparing the polarisation created on $|6\rangle\leftrightarrow|7\rangle$ or $|5\rangle\leftrightarrow|6\rangle$ after our protocol to their equilibrium thermal polarisation. We estimate that our DNP enhancement parameter is about $20$. 

We additionally confirm the population in $|6\rangle$ by monitoring the cooperativity $C_n$ via continuous-wave spectroscopy after each cycle (Fig.~\ref{fig:fig5}b). We measure $C_8/C_0 = \qty{2.8\pm0.1}{}$ after an 8-cycle protocol. This enhancement is smaller than the CPMG-based value because continuous-wave spectroscopy takes into account the contribution of weakly coupled and off-resonant spins which the CPMG spectroscopy disregards. These spins are located far from the resonator and are less likely to be efficiently repumped by the Purcell effect.

\section{Limiting mechanisms}

We next discuss the limitations of this DNP scheme. We first investigate how long this increase of population is sustained. Waiting a delay $\tau_D$ after running 8 cycles of DNP, we observe a decrease of the population due to the hyperfine diffusion process for which we extract an associated decay time $T_{\mathrm{hf}}= \qty{166 \pm 5}{\minute}$ (see Appendix~\ref{m:intrinsicT1}). We can also continuously monitor the population in $|6\rangle$ by repeatedly performing CPMG spectroscopy, with one measurement every 4 min, to assess the robustness of the population gain against manipulation of the electronic transition. We observe a similar decay, albeit on a slightly shorter timescale (green in Fig.~\ref{fig:fig5}c).

The observed enhancement of $\qty{4.6\pm0.2}{}$ falls short of both the theoretical maximum of $9$ and the value of $4.9$ predicted by a model accounting for finite pulse efficiencies and the competition between Purcell effect and intrinsic relaxation mechanisms (see Appendix~\ref{m:DNPnumerics}). To identify the reason for this discrepancy, we measure the polarisation build-up curve at each cycle $n$, fixing the wait times of the preceding $n-1$ cycles at $\tau_{W} = \qty{320}{\second}$ and varying $\tau_W^n$ (Fig.~\ref{fig:fig5}d). We fit each curve with the intrinsic electronic lifetimes and pulse efficiencies fixed at their independently measured values, leaving only the Purcell rate $\Gamma_{\mathrm{P},n}$ and the hyperfine diffusion as free parameters. The extracted $\Gamma_{\mathrm{P}n}$ decreases systematically with cycle number (see Fig.~\ref{fig:fig5}e), demonstrating that radiative relaxation becomes progressively less efficient. We attribute this slowdown to a many-body effect. As population accumulates in $|6\rangle$, the cooperativity of the $|6\rangle\leftrightarrow|15\rangle$ transition grows, pushing the system toward the strong-coupling regime where spin-spin interactions mediated by the resonator develop and compete with the Purcell decay. On average, this means the ensemble takes longer to relax radiatively, reducing the per-cycle DNP efficiency. 

This effect has been predicted for a few emitters coupled to a cavity~\cite{vyshnevyy2022,agarwal2024}. To capture the limitations we observe, we derive spin ensemble-resonator dynamics up to the third order using cumulant expansion and approximation. Keeping higher-order interaction terms, namely resonator-induced spin-spin flip-flop terms, provides two corrections: the spins themselves add a polarisation-dependent loss channel that broadens the resonator to $\tilde{\kappa} = \kappa + \kappa_{\mathrm{s}}$, and photons emitted by one spin can be reabsorbed by spins of similar frequency. Both effects contribute in a similar manner: in a dense ensemble, the Purcell relaxation no longer occurs at a fixed per-spin rate but is a polarisation-dependent decay at reduced rate. Incorporating this correction into a numerical model taking into account the frequency and coupling inhomogeneities (see Appendix~\ref{m:DNPnumerics}) allows us to model all build-up curves quantitatively without adjustable parameters and account for the observed enhancement. 

From this model, we can identify the limitations to our DNP mechanism. Suppressing cavity-mediated flip-flops however increases the polarisation enhancement to $4.9$, while perfect WURST pulses (or additionally the absence of hyperfine diffusion) would bring us to $6.7$ (or $8.1$), with the remaining discrepancy to $9$ being due to weakly coupled spins which are hard to reset radiatively. For this mechanism to be fully efficient, the Purcell effect should then largely dominate other processes. For instance, repeating this experiment with a low-impedance resonator and an inter-chip distance of 300 nm would increase the Purcell rate, allowing unit cooperativity to be reached in approximately 10 minutes even for a spin ensemble that is 20 times more dilute (see Appendix~\ref{m:DNPnumerics}). This indicates that Purcell DNP can also be employed in quantum memories operating in the $C=1$ regime and would enable lower concentration ensembles to be used.

\section{Conclusion}\label{sec:ccl}

We introduce a dynamic nuclear polarisation mechanism in which the dissipation driving the population transfer is not a material property but an externally controlled parameter. By exploiting the Purcell effect to selectively accelerate the decay of a single electronic transition in a Bi:$^{28}$Si electron–nuclear spin system, we implement an eight-cycle protocol that concentrates the ground-manifold population into a single state, reaching a population enhancement of \qty{1.79\pm0.04}{} per cycle and $\qty{4.6\pm0.2}{}$ for 8 cycles, to be compared with the ideal values of 2 and 9. Three mechanisms account for this difference: the finite ratio of Purcell to intrinsic relaxation rates, inversion infidelity, and cavity-mediated spin flip-flops.

The DNP enhancement parameter of our technique is lower than reported in solid state experiments (20 compared to 100-500 ~\cite{dai2025}) owing to our already sizeable equilibrium nuclear polarisation. This hyperpolarisation is nevertheless obtained without relying on a particular optical transition or on a fast intrinsic relaxation process. Bi:$^{28}$Si is well-suited to deterministic pumping into a target state as its transitions are well-resolved from one another, so that the superconducting resonator addresses and resets them one at a time. Beyond this platform, Purcell-DNP applies to solid-state spin systems with spectrally resolved electron–nuclear transitions and sufficient coupling to a microwave resonator, such as rare-earth ions, NV centers, and molecular spins. Its extension to nuclear spins with smaller coupling to the electron spin, such as the nuclear spin bath of the host material, or radicals, remains open. Succeeding would offer a route to increase NMR sensitivity in nanoscale and volume-limited samples. 

For quantum information applications, the most immediate impact is on spin-ensemble quantum memories: a cooperativity $C \sim 1$ in dilute spin systems allows coherent storage and retrieval of single microwave photons from a superconducting processor while preserving the coherence time of the ensemble~\cite{tyryshkin2012}. Our demonstration that Purcell-DNP increases the ensemble cooperativity by a factor $\qty{2.8 \pm 0.1}{}$ represents a  first step on this path.

\begin{acknowledgments}
We acknowledge funding from the Plan France 2030 through the project ANR-22-PETQ-0003.  T. L. acknowledges the support of France 2030 project QuanTEdu-France ANR-22-CMAS-0001. A. B. acknowledges support from the "Fondation CFM pour la Recherche". We acknowledge IARPA and Lincoln Labs for providing the Josephson traveling-wave parametric amplifier. We acknowledge the support of AIDAS joint laboratory, of Région Ile-de-France through the DIM SIRTEQ, and of the Agence Nationale de la Recherche under the Chaire Industrielle NASNIQ.
\end{acknowledgments}

\appendix
\section{Spin Hamiltonian: transition and matrix elements}
\label{m:hamiltonian}
In our experiment, the static magnetic field $\mathbf{B}_0$ is applied parallel to
the inductive strip used to couple the spin ensemble to the resonator and sets the
spin quantisation axis. The inductive strip generates an AC magnetic field
perpendicular to this quantisation axis, so that we only probe $i$--$j$
transitions whose $\bra{i}\hat{S}_x\ket{j}$ matrix element is non-zero. The
Hamiltonian of the bismuth donor spins is such that only transitions between
states with $\Delta m_{\mathrm{F}} = \pm 1$ possess non-zero $\hat{S}_x$ matrix
elements. The frequencies of these transitions are shown in Table~\ref{m:tableSpin}, along with their matrix elements. These values are obtained by diagonalizing the system Hamiltonian using the ratios $\gamma_n$ and $\gamma_e$ given in the main text (extracted from \cite{wolfowicz2013}). The hyperfine interaction strength and the exact magnetic field $B_0$ applied by the coil are extracted by adjusting the resulting transition frequencies to the 6 transitions frequencies we measure experimentally in Fig.~\ref{fig:fig2}a and Fig.~\ref{fig:fig3}c (indicated in bold). We obtain $A = \qty{1475.07\pm0.03}{\mega\hertz}$ and $B_0=\qty{65.371 \pm0.002}{\milli\tesla}$. The value of $A$ is very close ( $-\qty{0.1}{\mega\hertz}$) to what was measured in ~\cite{wolfowicz2013} for a bulk crystal, indicating there is only a minor residual strain in the spin host lattice~\cite{pla2018strain}.

\setlength\tabcolsep{3pt}
\begin{table*}[htbp]
\centering
\begin{tabular}{|c|c|c||c|c|c|}
\hline
\multicolumn{6}{|c|}{}\\[-6pt]
\multicolumn{6}{|c|}{Microwave transitions}\\[2pt]
\hline
\multicolumn{3}{|c||}{} & \multicolumn{3}{c|}{}\\[-6pt]
\multicolumn{3}{|c||}{$\Delta m_F=+1$} & \multicolumn{3}{c|}{$\Delta m_F=-1$}\\[2pt]
\hline
$i$-$j$ & $f_{i\text{-}j}$ (MHz) & $\langle i|S_x|j\rangle$ &
$i$-$j$ & $f_{i\text{-}j}$ (MHz) & $\langle i|S_x|j\rangle$ \\
\hline
1--20 & 9056.47 & 0.48 & 1--18 & 8754.06 & 0.05 \\
2--19 & 8753.15 & 0.45 & 2--17 & 8439.83 & 0.08 \\
3--18 & 8438.92 & 0.41 & 3--16 & 8113.43 & 0.12 \\
4--17 & 8112.52 & 0.37 & 4--15 & 7773.30 & 0.17 \\
5--16 & 7772.39 & 0.33 & $\bm{5}$--$\bm{14}$ & $\bm{7417.57}$ & $\bm{0.21}$ \\
$\bm{6}$--$\bm{15}$ & $\bm{7416.65}$ & $\bm{0.29}$ & 6--13 & 7043.84 & 0.26 \\
7--14 & 7042.93 & 0.23 & 7--12 & 6649.07 & 0.32 \\
8--13 & 6648.16 & 0.18 & 8--11 & 6229.23 & 0.38 \\
9--12 & 6228.32 & 0.11 & 9--10 & 5778.80 & 0.46 \\
\hline
\multicolumn{6}{c}{}\\[-6pt]
\hline
\multicolumn{6}{|c|}{}\\[-6pt]
\multicolumn{6}{|c|}{HF transitions}\\[2pt]
\hline
\multicolumn{3}{|c||}{} & \multicolumn{3}{c|}{}\\[-6pt]
\multicolumn{3}{|c||}{$F=4$} & \multicolumn{3}{c|}{$F=5$}\\[2pt]
\hline
$i$-$j$ & $f_{i\text{-}j}$ (MHz) & $\langle i|S_x|j\rangle$ &
$i$-$j$ & $f_{i\text{-}j}$ (MHz) & $\langle i|S_x|j\rangle$ \\
\hline
1--2  & 154.70 & 0.12 & 10--11 & 233.21 & 0.20 \\
2--3  & 160.44 & 0.16 & 11--12 & 216.31 & 0.25 \\
3--4  & 166.88 & 0.19 & 12--13 & 202.62 & 0.27 \\
4--5  & 174.16 & 0.21 & 13--14 & 191.24 & 0.28 \\
\textbf{5--6}  & \textbf{182.49} & \textbf{0.22} & \textbf{14--15} & \textbf{181.58} & \textbf{0.27} \\
\textbf{6--7}  & \textbf{192.15} & \textbf{0.22} & \textbf{15--16} & \textbf{173.25} & \textbf{0.26} \\
7--8  & 203.53 & 0.21 & 16--17 & 165.97 & 0.24 \\
8--9  & 217.22 & 0.17 & 17--18 & 159.53 & 0.21 \\
      &        &      & 18--19 & 153.79 & 0.18 \\
      &        &      & 19--20 & 148.62 & 0.13 \\
\hline
\end{tabular}
\caption{Frequencies and matrix elements for allowed $S_x$ transitions at
$B_0=\qty{65.371\pm0.002}{\milli\tesla}$. Bold entries are the transitions of relevance
in the main text.}
\label{m:tableSpin}
\end{table*}

\section{Device}
\label{m:device}

\begin{figure}[htbp]
    \centering
    \includegraphics[width=\linewidth]{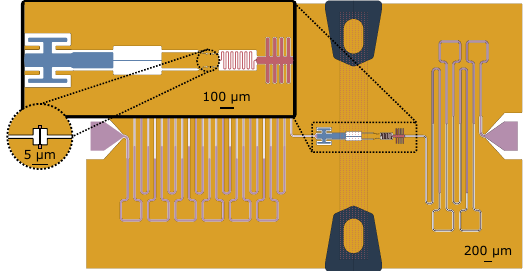}
    \caption{\label{figm:figDevice} Device layout. Overview of the NbTiN film (yellow) patterned on the silicon substrate, corresponding to the base chip before flip-chip bonding of the $^{28}$Si:Bi cover die. Two bonding pillars (dark blue) sit at the ends of the footprint where the Bi-doped silicon die is subsequently glued (darker yellow). Inset: zoom of the central region, showing the buffer interdigitated capacitor (red) in series with its \qty{4}{\micro\metre}-wide inductor (red meander), and the auxiliary \qty{800}{nm}-wide inductive microwire (blue) in series with its own interdigitated capacitor (blue, far left). Circle: additional zoom on the pair of $\qty{5}{\micro\metre}$-long, $\qty{250}{nm}$-wide NbTiN nanowires which provide the mutual inductance coupling resonators A and B. On the right (left) side of the central circuit, a 6(20)-section Bragg-mirror notch filter allows to combine high-quality factors lumped resonators at microwave frequencies while still passing DC bias and RF pump tones (light purple). The buffer and auxiliary resonators are addressed via the right port as the coupling to this port dominates over the right port for both modes. DC and hyperfine drives as discussed in this manuscript are applied to the left port.}
\end{figure}

The device used in this work is described in Ref.~\cite{wen2025}. It consists of a pair of 2D-lumped superconducting microwave resonators which are inductively coupled and respectively named "buffer" (B) and "auxiliary" (A) resonators. They are patterned out of a 50-nm NbTiN thin-film deposited on a silicon wafer, which allows them to be field-resilient. The auxiliary resonator is the high-quality factor resonator coupled to the spins discussed in the main text. Its frequency can be tuned over $\qty{80}{\mega\hertz}$ by passing a DC current through its inductance. The quality factor of the buffer resonator is about $1000$. We address the auxiliary resonator through this buffer resonator for continuous-wave spectroscopy and for driving the spins. CW spectroscopy of the spins is performed at low photon number ($\sim 10$). The measurement setup is described in the Supplemental Material~\cite{SM}.

The spins are introduced by flip-chipping a surface-implanted $^{28}$Si sample on top of the NbTiN chip.  The Bismuth atoms are implanted in this $^{28}$Si using various fluences and energies. A Stopping and Range of Ions in Matter (SRIM) simulation using these parameters as input yields the expected Bi concentration as a function of depth shown in the supplementary material of \cite{wen2025}, yielding an approximately constant concentration of Bismuth atoms $\qty{1e17}{\centi\meter\tothe{-3}}$ over a depth from $\qty{50}{\nano\meter}$ to $\qty{600}{\nano\meter}$. Only a fraction $\eta_{I}$ of these atoms are located in substitutional sites and effectively contribute as a donor spin. Depending on the annealing conditions, this fraction of atoms can vary. In our case, we expect this fraction to be 60\%~\cite{weis2012}.

The resonator includes a 800-nm-wide and 350-$\qty{}{\micro\meter}$-long inductive strip through which it couples to the flip-chip spin ensemble. Finite-element microwave simulations indicate the impedance of this resonator is $Z_0 = 160~\Omega$, resulting in a zero-point-fluctuation (ZPF) current $\delta i = \qty{33}{\nano\ampere}$ flowing through the inductor. We assume a homogeneous distribution of the current within the inductive strip to estimate the ZPF AC magnetic field $\delta \mathbf{B}(\mathbf{r})$ (a valid assumption as the width of the wire is near the London penetration depth of the NbTiN $\sim\qty{200}{\nano\meter}$).
The coupling strength $g$ between the resonator and the transition $\ket{i}\leftrightarrow\ket{j}$ of a bismuth spin located at $\mathbf{r}$ is $g(\mathbf{r})/(2\pi) = \gamma_{\mathrm{e}}\, |\bra{i} \hat{S}_x \ket{j}|\, |\delta \mathbf{B}(\mathbf{r})|$, assuming a Jaynes-Cummings interaction $\hat{H}_{\mathrm{int}} = \hbar g\,(\hat{a}\,\ket{j}\bra{i} + \hat{a}^{\dagger}\,\ket{i}\bra{j})$, where $\hat{a}$ is the annihilation operator of the resonator.
To describe the ensemble coupling distribution $\rho_g(g)$, one thus only needs to determine the location of the spins with respect to the inductive strip, determined by the implantation profile and the interchip distance $d$.

In the next paragraphs, we detail the experimental techniques used throughout the manuscript and their corresponding modelling that take into account the coupling and frequency inhomogeneity of the spin ensemble. We assume that there is no correlation between the spin location and its frequency, i.e. that the spin-resonator coupling and the spin frequency distributions $\rho_g$ and $\rho_\delta$ are uncorrelated. According to the CW spectroscopy shown in Fig.~\ref{fig:fig2}a the spin frequency distribution $\rho_\delta$ is reasonably estimated by a Lorentzian function of width $\Gamma_{6-15}$.

\section{Experimental setup}\label{m:setup}

\begin{figure}[ht]
    \centering
    \includegraphics[width=\columnwidth]{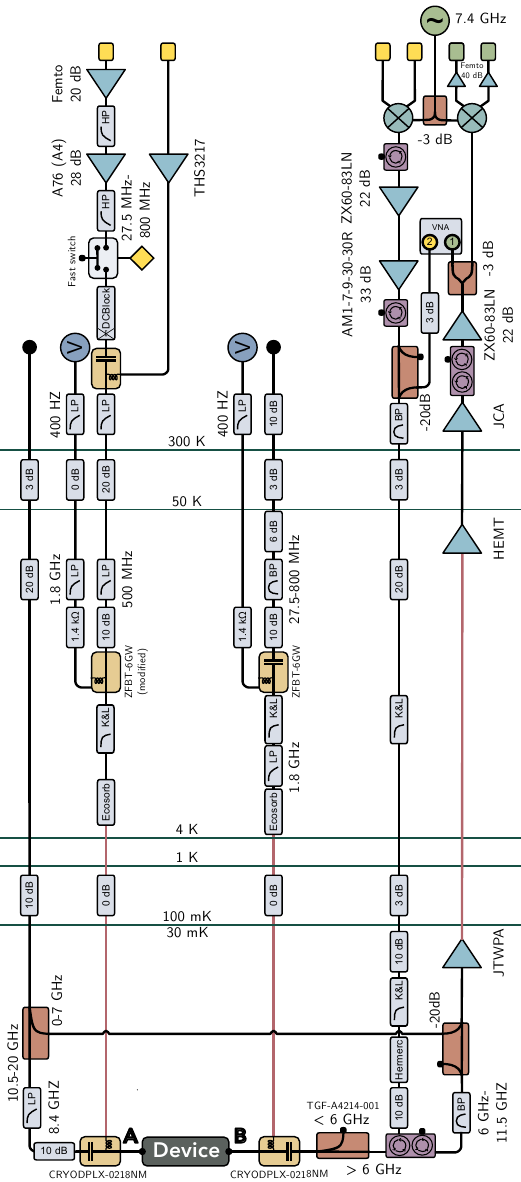}
    \caption{\label{apx:fig1} Schematic of the measurement setup. A color matched RF source is dedicated to each element.}
\end{figure}

The device is cooled down in a Bluefors LD250 cryostat, operating at a base temperature of~$\qty{25}{\milli\kelvin}$. It is placed inside a home-made 3D superconducting coil allowing to apply a static magnetic field~\cite{bahr2023}, itself protected by a cryoperm can. The cryogenic and room-temperature wiring diagram is shown in Fig.~\ref{apx:fig1}. 

To realize the experiments presented in the main text, we use 4 control signals (yellow squares) and two readout signals (green squares), generated and acquired by a Xilinx ZCU216 evaluation board controlled through the "QICK" firmware~\cite{stefanazzi2022} and Python library. We make use of 4 DACs and 2 ADCs of the firmware, which we operate at a sampling rate of $\qty{430}{\mega\sample\per\second}$ and $\qty{860}{\mega\sample\per\second}$. 
We also use two I/O pins cascaded with a GPIO level shifter to control two RF switches which prevent RF and microwave leakage from exciting the spins outside of the applied pulses. 

Two of the control DAC signals are used to generate the microwave drives used during the CPMG and electronic WURST pulses by upmodulating these signals using an IQ-mixer and a local oscillator provided by one of the channels of an AnaPico APUASYN20-4. The echoes generated subsequently by these pulses are amplified with a traveling-wave parametric amplifier kindly provided by Lincoln Labs, powered by a pump tone at $\qty{6.018}{\giga \Hz}$. The follow-up amplification is performed by a HEMT amplifier from Low Noise Factory at 4 K and by room-temperature amplifiers. The readout signal is then downconverted using an IQ mixer before further amplification and digitization by the ZCU216's ADCs.

The two remaining DACs are dedicated to the direct synthesis of a 200-MHz tone used to drive the hyperfine transitions and generates the nuclear WURST pulses, and to the generation of current pulses drives that allows to tune the auxiliary resonator frequency through the modulation of its kinetic inductance. For each of these DACs signals, we recombine the ZCU output differential DAC signals using a dedicated DC amplifier board (THS3217 from Texas Instruments). The two resulting control signals are then combined using a bias-tee before being further recombined at 4K with a static current applied using a Yokogawa voltage source associated to a cold bias resistor. The DC bias, DC pulse and RF drives are then fed to the device.

\section{Reset}
\label{m:reset}
The intrinsic electronic and hyperfine diffusion lifetimes being long ($\qty{24}{\minute}$ and $\qty{166}{\minute}$ respectively), the system takes a too long time to go back to its equilibrium value for practical experimental use. We therefore apply a reset sequence before each measurement by combining two mechanisms that act sequentially to form a reproducible initial state : emptying state $|15\rangle$  (or $|14\rangle$) via the modulated Purcell effect, and then saturating the hyperfine levels to equalize the populations across this manifold.
 
First, the resonator is tuned onto the electronic transition and its frequency is
modulated, so that Purcell relaxation empties $|15\rangle$ into $|6\rangle$ (or $\ket{14} $ into $\ket{5}$,
depending on the measurement); the effective Purcell time is
$T_1^{\mathrm{eff}} = \qty{41(1)}{\second}$ (see Fig.~\ref{fig:fig2}b). This reset lasts between $\numrange{2}{4}\,\si{\minute}$, de exciting $|15\rangle$ down to $\numrange{5}{0.3}\,\si{\percent}$ of its initial population.

Then, the ground-manifold populations are equalised by saturating the hyperfine transitions addressed by the measurement sequence. A \qty{200}{\micro\second} band-limited noise waveform, spanning $\qty{\pm 0.5}{\mega\hertz}$ about each target frequency, is played $\qty{100}{}$ times per second at each of these frequencies. Being incoherent and $\qty{3.8}{}$ times broader than the inhomogeneous linewidth, it equalizes the population rather than inverting them. Its repetition rate is fast compared to both the Purcell decay and the nuclear diffusion, so the ground populations are only affected by the scrambling process. We therefore take the nine ground states as equally populated at the start of each sequence.

This procedure is for instance used for the fidelity measurements of Sec. Appendix~\ref{m:fidelities}, with a reset time of $\tau_{\mathrm{reset}} = \qty{120}{\second}$ and the hyperfine frequency list restricted to $\ket{6}\leftrightarrow\ket{7}$.

\section{CPMG spectroscopy}
\label{m:CPMG}

To better understand the dynamics probed via the CPMG echo, we would like to determine the contribution of each spin to the signal depending on their detuning $\delta$ and coupling strength $g$ to the resonator. We do this using the methodology presented by Hurlimann and Griffin~\cite{hurlimann2000} to determine the evolution of the ensemble magnetization during the CPMG sequence. 

We assume the CPMG pulses are applied on resonance with the resonator, so that their amplitude $\beta$ corresponds to loading $n_\beta= 4 \kappa_{\mathrm{c}} \beta^2/(\hbar \omega_R \kappa^2)$ photons into the resonator, where $\kappa_{\mathrm{c}}/2 \pi = \qty{15}{kHz}$ is the coupling rate of the resonator. We consider that the durations $t_{\pi/2}$ and $t_\pi$ of the $\pi/2$ and $\pi$  pulses are sufficiently long (respectively 6 and $\qty{12}{\micro\second}$) so that we can neglect the cavity response and consider that this intra-resonator population is reached instantaneously for each pulse. For a given pulse, each spin magnetization vector $\vec{m}$ thus undergoes a rotation of nutation frequency $\Omega = \sqrt{\delta^2+4g^2n_\beta}$ around an axis $\vec{n} = \frac{2g\sqrt{n_\beta}}{\Omega} [\cos{\phi}  \vec{x} + \sin{\phi}  \vec{y}] + \frac{\delta}{\Omega} \vec{z}$, where $\phi$ is the phase of the drive ($\qty{0}{\degree}$ for the $\pi/2$ pulse, and $\qty{90}{\degree}$ for the 100 successive $\pi$ CPMG pulses), and $\vec{x}$, $\vec{y}$, $\vec{z}$ are the geometrical axes of the spin Bloch sphere representation. Between the pulses, the magnetization evolves around the $z$-axis, with $[m_x+ i m_y](t+dt) = \exp(i \delta dt)[m_x+ i m_y](t)$. 

We can decompose the CPMG sequence into block rotations, a first block $\mathcal{R}_{\pi/2}$ comprising an initial rotation due to the $\pi/2$ pulse which creates the magnetization followed by a wait time $\tau/2$, followed by a refocussing block $\mathcal{R}_{\pi}$ consisting in a wait time $\tau/2$, a $\pi$ pulse and a wait time $\tau/2$. The remainder of the sequence is modelled by an additional block consisting of a wait time $\tau/2$, and $N-1$ refocussing blocks $\mathcal{R}_{\pi}$ corresponding to the additional refocussing actions of the CPMG sequence. Each block can be described by an overall rotation whose angle and axis can be calculated by applying composite rotations for each possible value of $g$ and $\delta$. The magnetization of each spin during the CPMG sequence can thus be tracked by applying these rotations, though this treatment neglects any decoherence occurring during the sequence. Alternatively, the asymptotic magnetization for each $g$ and $\delta$ is given as $\vec{m} = \vec{n}_{\mathcal{R}_{\pi}} (\vec{n}_{\mathcal{R}_{\pi}} \cdot \mathcal{R}_{\pi/2}(\vec{z}))$~\cite{hurlimann2000}. We make the approximation that measuring and averaging the first 100 echoes through heterodyne detection is analogous to monitoring this asymptotic magnetization. The echo is thus proportional to the following signal~\cite{bienfait2015}: 
\begin{equation}
\langle \hat{a}_{\mathrm{out}} \rangle = \sqrt{\kappa_c} \sum_\delta \sum_g \frac{2 i g}{\kappa + 2 i \delta} (m_x - i m_y) \rho_g(g)\rho_\delta(\delta)
\label{aout}
\end{equation}

\begin{figure}[htbp]
    \centering
    \includegraphics[width=\linewidth]{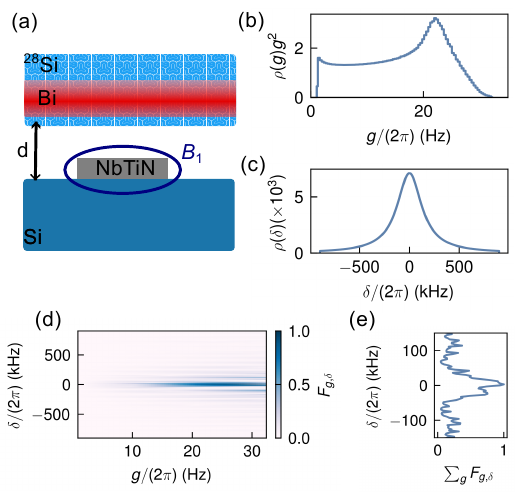}
    \caption{(a) A $^{28}$Si chip with a surface implantation of Bi atoms is overlaid on a 800-nm wide inductor with an inter-chip distance $d$. (b) Spin-resonator coupling distribution resulting from the integration of the inductance in a $160~\Omega$ resonator, for $d=\qty{1100}{\nano\meter}$, and for the $6-15$ transition, containing $81$ bins. (c)  Spin frequency Lorentzian distribution, containing 601 bins. (d)  Simulated echo response function as described in the main text. (e) Simulated CPMG average echo spectral bandwidth when integrated over all coupling constants, showing an echo bandwidth of $\sim \qty{40}{\kilo\hertz}$.}
    \label{figm:rhog}
\end{figure}

If one divides the ensemble of spins into families of identical coupling strength $g$ and detuning $\delta$, containing $N_{g,\delta}$ spins (a number set according to the $g$ and $\delta$ distributions described earlier), it is possible to write Eq.(\ref{aout}) as $\langle \hat{a}_{\mathrm{out}} \rangle = \sum_g \sum_\delta F_{g,\delta} N_{g,\delta}$ where $F_{g,\delta}$ expresses the contribution of each spin family to the signal. This matrix $F_{g,\delta}$ is plotted in Fig.~\ref{figm:rhog}.  We can observe that only the spins lying in a $\sim\qty{40}{\kilo\hertz}$ bandwidth of the drive participate to the CPMG signal.

\section{Intrinsic relaxation mechanisms}
\label{m:intrinsicT1}

\subsection{Modelling}
Ideally, to estimate precisely the DNP effect, we would like to know all possible decay paths for any transition and their associated decay rates. However, measurements of intrinsic lifetimes of Bismuth donors at low temperature are few, and the long timescale associated with these lifetimes ($\sim 20$ min) makes it difficult to run a complete analysis. A general map of these decay paths is also missing. For silicon donors, Feher and Gere~\cite{feher1959} summarize four decay mechanisms: direct spin relaxation, connecting states of same quantum number $m_{\mathrm{S}}$ (corresponding to $\Delta m_{\mathrm{F}}=-1$), a nuclear relaxation process (e.g. $|15\rangle$ relaxing to $|14\rangle$ and $|16\rangle$, or $|6\rangle$ relaxing to $|5\rangle$ and $|7\rangle$ ),  cross-spin relaxation involving an electro-nuclear flip ($\Delta m_{\mathrm{F}}=0$ transitions, e.g. $|15\rangle$ relaxing to $|5\rangle$) and finally electron-electron flip-flops (near-energy conserving flip-flops on two $\Delta m_{\mathrm{F}}=1$ transitions, e.g. deexciting $|15\rangle$ to $|5\rangle$ at the price of exciting $|6\rangle$ to $|14\rangle$). At the low-field and low-temperature we operate at we expect the latter two mechanisms to be strongly suppressed. In addition, since in our low-field limit, transitions between highly-mixed states are allowed ($\Delta m_{\mathrm{F}}=+1$), we make the assumption that direct spin relaxation will also impact these transitions albeit at a different rate, as well as the nuclear-like transitions. In the absence of better knowledge, we make the assumption that we can distinguish two electronic relaxation processes ($\Delta m_{\mathrm{F}}=+1$,$\Delta m_{\mathrm{F}}=-1$), of rate independent of the initial $m_F$ state. We thus distinguish nine direct electronic relaxations of identical rate $\Gamma_e^{-1}$ :
\begin{equation}
    |F=5, m_{\mathrm{F}} \rangle \xrightarrow{\Gamma_e^{-1}} |F=4, m_{\mathrm{F}}-1 \rangle,
\end{equation}
e.g. deexciting the populations of states $|20\rangle$, $|19\rangle$,...,$|12\rangle$ towards states $|1\rangle$, $|2\rangle$,...,$|9\rangle$. We also consider nine cross electronic relaxation of identical rate $\Gamma_e^{+1}$:
\begin{equation}
    |F=5, m_{\mathrm{F}} \rangle \xrightarrow{\Gamma_e^{+1}} |F=4, m_{\mathrm{F}}+1 \rangle,
\end{equation}
e.g. deexciting the populations of states $|18\rangle$, $|17\rangle$,...,$|10\rangle$ towards states $|1\rangle$, $|2\rangle$,...,$|9\rangle$.
As the smallest electronic-like transition is well into the gigahertz regime and as we operate at $T=\qty{25}{\milli\kelvin}$, we neglect any thermal excitation from the $F=4$ manifold to the $F=5$ manifold.

Last, considering the temperature of the experiment and the very similar hyperfine frequencies between neighbouring nuclear states, we consider a simplified version of hyperfine diffusion governing the nuclear population dynamics through:
\begin{equation}
    \frac{d}{dt} p_i = \Gamma_{\mathrm{N}} (p_{i+1} + p_{i-1} - 2p_i),
\end{equation}
with a state-independent diffusion rate $\Gamma_{\mathrm{N}}$.

\subsection{Fits}
\begin{figure}[htbp]
    \centering
    \includegraphics[width=\linewidth]{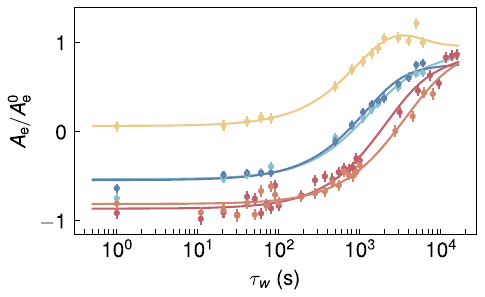}
    \caption{\label{figm:figT1int}Intrinsic lifetimes measured on the
$\ket{6}\leftrightarrow\ket{15}$ transition after inverting
$\ket{6}\leftrightarrow\ket{15}$ (blue), after inverting
$\ket{5}\leftrightarrow\ket{14}$ and reading that transition (cyan), after
storing the population in $\ket{20}$ (red) or in $\ket{10}$ (orange) during the
wait, and after inverting $\ket{6}\leftrightarrow\ket{15}$ and
$\ket{5}\leftrightarrow\ket{14}$ simultaneously (yellow). Solid lines are a joint fit described in the text; each dataset carries its own amplitude scale factor.}
\end{figure}

To extract the values of these three parameters, $\Gamma_{\mathrm{N}}$, $\Gamma_{\mathrm{e}}^{+1}$ and $\Gamma_{\mathrm{e}}^{-1}$, we make use of six datasets. First, from the DNP hyperpolarisation decay (Fig.~\ref{fig:fig5}b, blue curve), we adjust $\Gamma_{\mathrm{N}}$ alone, as the electronic processes play no role there. We find $1/\Gamma_{\mathrm{N}} = \SI{166(5)}{\minute}$. We then run five different experiments:

\begin{itemize}
\item Invert the population of $\ket{6}$ to $\ket{15}$ with an electronic WURST pulse, and monitor the polarisation of the $\ket{6}\leftrightarrow\ket{15}$ transition by a CPMG measurement after a wait time $\tau$. The resonator is detuned by $\qty{1.7}{\mega\hertz}$ during this measurement. This dataset appears in Fig.~\ref{figm:figT1int} (blue curve) and in Fig.~\ref{fig:fig2}b (blue curve).
\item Same protocol on $\ket{5}\leftrightarrow\ket{14}$, again with the resonator
far detuned (Fig.~A2, cyan curve).
\item Invert both $\ket{5}\to\ket{14}$ and $\ket{6}\to\ket{15}$ with two consecutive electronic WURST pulses, and monitor $\ket{6}\leftrightarrow\ket{15}$ (Fig.~\ref{figm:figT1int}, yellow curve). This dataset confirms that no flip-flop process takes place.
\item Invert $\ket{6}$ to $\ket{15}$, then shuffle the population to $\ket{20}$ with five successive nuclear WURST pulses. After a wait time $\tau$, shuffle the remaining population of $\ket{20}$ back to $\ket{15}$ and monitor $\ket{6}\leftrightarrow\ket{15}$ (Fig.~\ref{figm:figT1int}, red curve). This isolates the $\Gamma_{\mathrm{e}}^{-1}$ mechanism.
\item Same protocol via $\ket{10}$ instead of $\ket{20}$ (Fig.~\ref{figm:figT1int}, orange curve), which isolates $\Gamma_{\mathrm{e}}^{+1}$. \end{itemize}

We model the data by solving the population dynamics
\begin{equation}
  \dot{P} = (G_{\mathrm{e}}^{+1} + G_{\mathrm{e}}^{-1} + G_{\mathrm{n}})\,P,
  \label{eq:ode}
\end{equation}
where $P$ is a population vector and each matrix $G_x$ encodes one process according to its selection rules. The five curves are fitted jointly, each with an initial population matching its experimental sequence. Free parameters are the two electronic rates, one electronic inversion efficiency per sequence (the curves were acquired with different WURST settings), a single nuclear inversion efficiency $\eta_{\mathrm{n}}$ common to all curves, and one amplitude scale factor per dataset. Both nuclear shuffling steps are included, so $\eta_{\mathrm{n}}$ enters as $\eta_{\mathrm{n}}^{10}$ in the $\ket{20}$ and $\ket{10}$ sequences.

The fit gives $\eta_{\mathrm{n}} = \qty{98.6 \pm 0.4}{\percent}$, $\Gamma_{\mathrm{e}}^{+1} = \num{5.0(5)e-4}\,\si{\per\second}$ and $\Gamma_{\mathrm{e}}^{-1} = \num{1.9(5)e-4}\,\si{\per\second}$. For the excited state of interest here, $\ket{15}$, the resulting lifetime is $T_{1} ^{\mathrm{int}} = \SI{24(2)}{\minute}$.

\section{WURST inversion pulses}
\label{m:WURST}
We use WURST-20 pulses~\cite{odell2013}, corresponding to microwave pulses $A(t) \exp(i \omega_{\mathrm{d}}(t) t)$ whose amplitude $A$ and drive frequency $\omega_{\mathrm{d}}$ evolve as a function of $t$ as:
\begin{align}
A(t) &= 1-|\sin(\pi*\frac{t}{t_{\mathrm{W}}})|^{20} \times \textrm{rect}(\frac{t}{t_{\mathrm{W}}}) \\
\omega_{\mathrm{d}} - \bar{\omega_{\mathrm{d}}} &= 2\delta_W \frac{t}{t_{\mathrm{W}}} \times  \textrm{rect}(\frac{t}{t_{\mathrm{W}}})
\end{align}
where $t_{\mathrm{W}}$ is the duration of the pulse, set to $\qty{300}{\micro\second}$ for all pulses in our experiment.

\subsection{Experimental fidelities}
\label{m:fidelities}

The electronic and the nuclear inversion efficiencies are extracted from a single model of the complete measurement sequence, fitted jointly to the two
datasets of Fig.~\ref{fig:fig3}b and~\ref{fig:fig3}d. This model is elaborated by modelling each step of the measurement sequence of the entire datasets. First, a reset step (see Appendix~\ref{m:reset}) equalises the ground nuclear populations, which
we model by setting $p_i = 1$ for $i = 1,\dots,9$ and $p_i = 0$ otherwise. Each
of the $\qty{125}{}$ measurement points of a curve - one for every four WURST
pulses - is acquired sequentially, and we model one repetition as:
\begin{itemize}
  \item Apply one electronic WURST pulse on $\ket{6}\leftrightarrow\ket{15}$,
  modelled as a swap of the two populations with efficiency $\eta_{\mathrm{e}}$:
  $p_{6} \rightarrow \eta_{\mathrm{e}}\,p_{15} + (1-\eta_{\mathrm{e}})\,p_{6}$
  and
  $p_{15} \rightarrow \eta_{\mathrm{e}}\,p_{6} + (1-\eta_{\mathrm{e}})\,p_{15}$,
  so that the population difference is contracted by
  $2\eta_{\mathrm{e}}-1$ and changes sign at each pulse.
  \item Apply $N$ nuclear WURST pulses on $\ket{6}\leftrightarrow\ket{7}$, each
  modelled as a swap of $p_6$ and $p_7$ with efficiency $\eta_{\mathrm{n}}$.
  
  \item Measure the echo on $\ket{6}\leftrightarrow\ket{15}$ by a CPMG sequence,
  with $A_{\mathrm{e}} \propto p_{15} - p_{6}$. The measurement also saturates
  the transition, which we model by
  $p_{6},\,p_{15} \rightarrow (p_{6}+p_{15})/2$.

  \item Apply the reset sequence of Sec. Appendix~\ref{m:reset}: Purcell decay from $\ket{15}$ to $\ket{6}$ with characteristic time $\qty{41}{\second}$, followed with saturation of the $\ket{6}\leftrightarrow\ket{7}$ transition. We thus let evolve the populations according to Eq.~\ref{eq:ode} with the intrinsic rates of Sec. Appendix~\ref{m:intrinsicT1}, adding the Purcell term for the $\qty{120}{\second}$ wait time , and project $p_{6},\,p_{7} \rightarrow (p_{6}+p_{7})/2$ at each integration step.
\end{itemize}

When testing the nuclear inversion, the polarisation in Fig.~\ref{fig:fig3}d converges to $\qty{0.24}{}$ rather than $\qty{0.5}{}$. We attribute this to the fact that for this characterisation the hyperfine reset is applied only to $\ket{6}\leftrightarrow\ket{7}$, so that population slowly leaks out of $\ket{6}$ and $\ket{7}$ into the remaining ground states. We capture it by adding a leakage at a rate $\ell$ per repetition.

Using this step-by-step modelling, a fit with only the inversion fidelities and
the leakage rate as free parameters gives
$\eta_{\mathrm{e}} = \qty{0.9996 \pm 0.0001}{}$,
$\eta_{\mathrm{n}} = \qty{0.991 \pm 0.001}{}$ and $\ell = \qty{0.003 \pm 0.001}{}$ per repetition; the model reproduces the observed floor ($\qty{0.25}{}$ against $\qty{0.24}{}$).

\subsection{Modelling the inversion fidelity on the spin ensemble}
\begin{figure}[htbp]
    \centering
    \includegraphics[width=\linewidth]{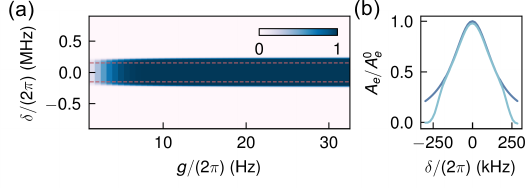}
    \caption{\label{figm:wurst} (a) Population inversion ratio of the $|6\rangle-|15\rangle$ transition created by $\pm \qty{250}{\kilo\hertz}$-range WURST pulse depending on the spin coupling $g$ and detuning $\delta$. Dashed lines represent the FWHM of the spin ensemble. (b) Simulated echo spectroscopy after a WURST inversion (dark blue: before, light blue : after, with the echo signed amplitude inverted for easier comparison), reproducing qualitatively the experimental observation of Fig.~\ref{fig:fig3}a.}
\end{figure}

To better understand the spin-dependent DNP limitation, we are also interested in modelling the inversion fidelity for a given spin with coupling strength $g$ and $\delta$. Using Dynamiqs~\cite{guilmin2025}, we thus establish inversion maps for any $\delta$ and $g$ values. We model the action of the WURST pulse on the spin by its evolution under the Hamiltonian $H/\hbar =  (\delta-\delta_W(t)) \sigma_z/2 + 2 g A_W(t) \sigma_x$, where $\delta_W$ and $A_W$ are the WURST control functions. We only consider a direct drive Hamiltonian and do not include the resonator dynamics in this modelling since the WURST drive is always applied resonantly (through kinetic modulation of the resonator frequency) and its length far exceeds the resonator loading time. We adjust the amplitude $A_W$ by comparison to the power necessary to obtain a maximum CPMG signal both experimentally and numerically. The simulated inversion map for the $g$ and $\delta$ range discussed in M. Appendix~\ref{m:device} is shown in Fig.~\ref{figm:wurst}. In particular, it shows that the inversion is more than $90\%$-efficient on a range $\pm\qty{220}{\kilo\hertz}$ for spins with coupling higher than $\qty{5}{\hertz}$.

\section{Engineered Purcell decay}
\label{m:wigglyPurcell}

\subsection{Dynamic resonator modulation}
A simple reset protocol for the spin ensemble would be to place the resonator at resonance with the central spin frequency, and let the Purcell effect take place to relax the spin ensemble. However, the Purcell effect, given by $\Gamma_\mathrm{P}=\kappa \frac{g^2}{\Delta^2+\kappa^2 / 4}$, is strongly dependent on the spin-resonator coupling and the spin-resonator detuning. Since, in our case, the spin-ensemble linewidth exceeds the resonator linewidth by a factor of $1.5$, the Purcell effect acts with limited efficiency for spins on the edge of the frequency distribution.

\begin{figure}[htbp]
    \centering
    \includegraphics[width=\linewidth]{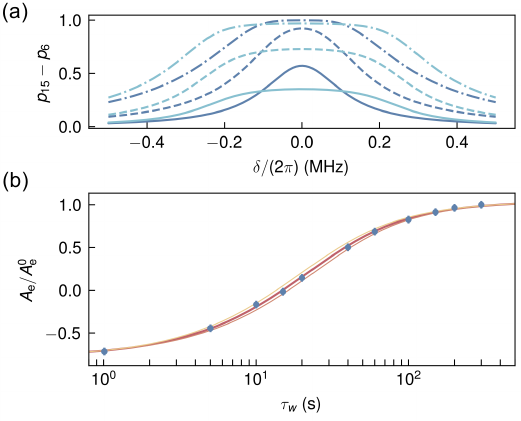}
    \caption{(a) Effect of modulation on the effective Purcell-induced polarisation. We consider an excited spin of coupling rate $g/(2\pi) = \qty{15}{Hz}$ placed in a resonator of decay rate $\kappa /(2\pi)=  \qty{200}{\kilo\hertz}$ whose frequency remains constant (dark blue curves) or is modulated (light blue curves) over a $\pm\qty{150}{\kilo\hertz}$ range at a fast rate compared to the Purcell rate. The resonator de-excites the spin with a resulting polarisation profile dependent on the spin-resonator detuning $\delta$ after a time $T$ of $30$, $90$ or $\qty{240}{second}$ (respectively solid, dashed, dot-dashed lines), exhibiting a better polarisation of the spin ensemble on the edges in presence of modulation at the price of a slower polarisation time. (b) We simulate the $T_1$ relaxation shown in Fig.~\ref{fig:fig2}b (in absence of modulation) by modelling successively the WURST inversion, the Purcell relaxation and the CPMG measurement, taking into account their dependence on the coupling and frequency distributions of the spin ensemble that we determine for different inter chip distance $d$ as the only free fit parameters. We find that the best fit to the experiment (blue markers) corresponds to $d=\qty{1.1}{\micro\meter}$ (red solid lines), with a $100$-nm uncertainty (orange and yellow curves correspond to $d=\qty{1.2}{\micro\meter}$ and $d=\qty{1.0}{\micro\meter}$).}
    \label{figm:polaT1}
\end{figure}

To implement an efficient reset mechanism for all spins, we chose to modulate the resonator frequency in a saw-tooth pattern (with a period of $\qty{18}{\micro\second}$). The resonator is thus modulated over the entire spin linewidth (the modulation range is estimated to be $\pm\qty{150}{\kilo\hertz}$) and enables the Purcell effect to take place over a wider range than the natural resonator linewidth, see Fig.~\ref{figm:polaT1}. It thus allows to reset the edge spins more efficiently, at the price of a longer overall relaxation time due to the reduced resonator dwell time.

\subsection{Estimation of inter chip spacing}
\label{m:T1modeling}
Finally, to estimate the spacing between the spins chip and the coupling inductor, we realize a full simulation of the inversion recovery sequence shown in Fig.~\ref{fig:fig1}b, with and without DC-assisted recovery. For a given distance $d$, we compute the $g$-distribution as shown in Fig.~\ref{figm:rhog}. We then calculate the CPMG response map, and adjust the corresponding power necessary to reproduce the Rabi oscillations we observe. Using this calibration, we then calculate the inversion map $\mathcal{I}$ for a WURST pulse (see Fig.~\ref{figm:wurst}). The experimental sequence is then numerically modelled by starting from an effective polarisation given by $p_0(g,\delta) = 1 - \exp( -\tilde{\Gamma}_p(g,\delta) T_R)$, applying the WURST inversion matrix, the $g-\delta$ dependent Purcell relaxation, and using the resulting polarising map to weight the CPMG echo response function. We plot the resulting numerical signal in Fig.~\ref{figm:polaT1} for various inter chip distances $d$. The most reasonable match seems to be around $d = \qty{1.1}{\micro\meter}$, justifying our use of this number throughout the text.

\section{Dynamical nuclear polarisation}

\subsection{DNP pulse sequence}
\label{m:DNPscheme}
\begin{figure}[htbp]
    \centering
    \includegraphics{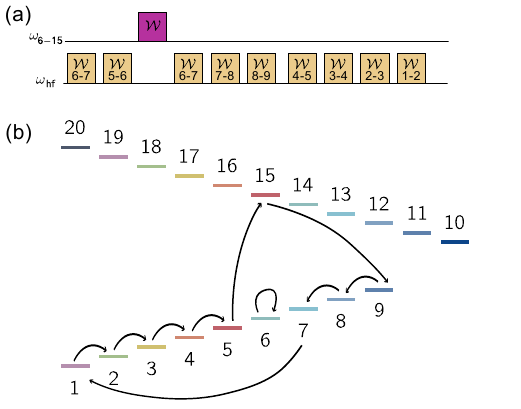}
    \caption{\label{figm:DNPscheme} a) Nuclear pumping sequence and b) idealized population transfer. As $|5\rangle$ is effectively excited to $|15\rangle$, its population gets added to state $|6\rangle$ through radiative relaxation, leaving $|5\rangle$ empty until additional hyperfine pulses transfer population to it. Population of $|6\rangle$ is stationary and grows at each cycle. Via the train of WURST pulses, state $|15\rangle$ which is effectively empty at the beginning of each cycle creates a hole in the manifold, creating 8 empty states after the completion of 8 cycles. An execution of the DNP scheme without imperfections  empties the populations in $|9\rangle$ for one cycle, $|8\rangle$ with a second cycle, $|7\rangle$ with a third, and then $|1\rangle \dots |5\rangle$ for successive pumping cycles. }
\end{figure}

The complete sequence of pulses used to realize the DNP scheme is shown in Fig.~\ref{figm:DNPscheme}. For each cycle, the same sequence of pulses is played.

\subsection{Fit of DNP dynamics leaving radiative and hyperfine relaxation as a free parameter}
\label{m:DNPODE}
To extract the effective Purcell relaxation rate induced by the resonator at each DNP cycle, we take as fixed parameters the intrinsic decay paths determined in Appendix~\ref{m:intrinsicT1} except for the hyperfine decay. To model the evolution of each cycle $n$, we follow the evolution of the population vector $\mathcal{P}$ according to Eq.\ref{eq:ode} to which we add a relaxation process of rate $\Gamma_i$ between $|15\rangle$ to $|6\rangle$. We adjust this rate by a global numerical factor so that $p_{15} - p_{6}$ matches the observed echo variation $A_e/A_e^0$. The initial population at each cycle is taken as the population of cycle $n-1$ at time $\tau_{n-1}=\qty{320}{\second}$ convolved with the application of the WURST inversion pulses of the DNP protocol taking into account their finite efficiencies (see Fig.~\ref{figm:DNPscheme}).

\section{Breakdown of the Purcell effect in a collective regime}

\subsection{Theory}
We introduce the limits of the Purcell effect when considering its application to an ensemble of spins. We wish to describe the dynamics of our system up to a third order cumulant expansion. In the reference frame of the auxiliary resonator (with annihilation and creation operators $a$ and $a^\dagger$), the Hamiltonian of the system reads:
\begin{equation}
    H = \sum_j \hbar\delta_j \frac{\sigma_z^j}{2} + \sum_j \hbar g_j (a^\dagger \sigma_-^j + a \sigma_+^j),
\end{equation}
where each spin $j$ is described through its Pauli operators, and its detuning $\delta_j$ and coupling $g_j$ to the cavity. We assume that the spin frequency distribution follows a Lorentzian distribution of width $\Gamma$ and centered on the resonator frequency. We assume the only imperfections of our system are the homogenous broadening of each spin $\gamma$ and the cavity decay rate $\kappa$, i.e. we neglect other relaxation mechanism for the spins. We derive the dynamics of the system in absence of a drive, so that the dynamics of the system are given by:
\begin{equation}
\dot{\rho} = -\frac{i}{\hbar} \left[ H, \rho\right] + \mathcal{L}[\sqrt{\kappa} a]\rho + \sum_j  \mathcal{L}[\sqrt{\gamma/2} \sigma_z^j]\rho.
\end{equation}

The evolution of the polarisation of a spin $j$ is given by $p_j = -\langle \sigma_z^j\rangle$, whose dynamics are given by:
\begin{equation}
\label{sigma_z_brut_de_pomme}
\frac{d}{dt} \langle \sigma_z^j\rangle = -2 i g_j (\langle a \sigma_+^j\rangle - \langle a^\dagger \sigma_-^j\rangle).
\end{equation}
The dynamics of the term $\langle a \sigma_+^j\rangle$ is given by:
\begin{equation}
    \begin{split}
   \frac{d}{dt} \langle a \sigma_+^j\rangle &= - (\kappa/2 - i \delta_j) \langle a \sigma_+^j\rangle\\
   &\quad - i \frac{g_j}{2} (1+\langle\sigma_z^j\rangle+2\langle \sigma_z^j a^\dagger a \rangle) \\
   &\quad- \sum_{k \neq j} i g_k \langle \sigma_-^k \sigma_+^j \rangle.
    \end{split}
\end{equation}
A third-order cumulant can be approximated to $\langle o p q \rangle = \langle o\rangle \langle p q \rangle + \langle p\rangle \langle o q \rangle+\langle q\rangle \langle o p \rangle - 2 \langle o\rangle \langle p \rangle \langle q \rangle$~\cite{kubo1962}. In the absence of a drive, the term $\langle \sigma_z^i a^\dagger a \rangle$ can thus be simplified into $\langle \sigma_z^j\rangle \langle a^\dagger a \rangle = \langle \sigma_z^j\rangle \bar{n}$, where $\bar{n}$ is the thermal occupancy of the resonator. We make the assumption that this term evolves adiabatically  ($\frac{d}{dt} \langle a \sigma_+^j\rangle =0$) as the Purcell effect is a very slow effect, much slower than any other dynamics in the system. We also assume $\gamma<\kappa$. We thus find that:
\begin{equation}
  \label{asigma}
  \begin{split}
    \langle a\sigma_+^j\rangle
    &= -\frac{i g_j/2}{\kappa/2 - i\delta_j}
       \left[1 + (1+2\bar{n})\langle\sigma_z^j\rangle\right] \\
    &\quad - \sum_{k\neq j} \frac{i g_k}{\kappa/2 - i\delta_j}
       \langle\sigma_-^k\sigma_+^j\rangle .
  \end{split}
\end{equation}
Ignoring the flip-flop term on the right-hand side, re-injecting this expression in Eq.~\ref{sigma_z_brut_de_pomme} and evaluating $\langle a^\dagger \sigma_-^j\rangle$ leads us to recover the Purcell effect:
\begin{equation}
     \frac{d}{dt} \langle \sigma_z^j\rangle = - \frac{g_j^2 \kappa}{(\kappa/2)^2 + \delta_j^2} (1+ (1+2 \bar{n}) \langle \sigma_z^j\rangle),
\end{equation}
where we recognize $\Gamma_P^j = \frac{g_j^2 \kappa}{(\kappa/2)^2 + \delta_j^2}$. We will now show that in presence of many spins, as in our experiment, the flip-flop term that we have discarded can actually play a role in the dynamics. Deriving the evolution of $\langle  \sigma_-^k \sigma_+^j  \rangle $ and performing again the assumption of adiabaticity leads us to:
\begin{equation}
  \label{eq:paircoh}
  \begin{split}
    \langle\sigma_-^k\sigma_+^j\rangle
    &= \frac{i g_k}{2\gamma - i\delta_{jk}}
       \langle\sigma_z^k\rangle\langle a\sigma_+^j\rangle \\
    &\quad - \frac{i g_j}{2\gamma - i\delta_{jk}}
       \langle a^\dagger\sigma_-^k\rangle\langle\sigma_z^j\rangle ,
  \end{split}
\end{equation}
where $\delta_{jk}=\delta_j-\delta_k$. Injecting this expressing in Eq.~\ref{asigma}, we obtain:
\begin{equation}
  \label{asigma_ff}
  \begin{split}
    \left(\frac{\kappa}{2} - i\delta_j\right)&\langle a\sigma_+^j\rangle
    = -i\frac{g_j}{2}
      \left[1 + (1+2\bar{n})\langle\sigma_z^j\rangle\right] \\
    &\quad + \left(\sum_{k\neq j}
        \frac{g_k^2}{2\gamma - i\delta_{jk}}\langle\sigma_z^k\rangle\right)
        \langle a\sigma_+^j\rangle \\
    &\quad - g_j \langle\sigma_z^j\rangle
      \left(\sum_{k\neq j} \frac{g_k}{2\gamma - i\delta_{jk}}
        \langle a^\dagger\sigma_-^k\rangle\right) .
  \end{split}
\end{equation}

The second term has the dimensions of a rate and can be absorbed into the
left-hand side, defining a resonator response dressed by the spins. One one side, the resonator decay rate becomes dependent on the spin polarization $\tilde{\kappa}_j = \kappa + \tilde{\kappa}_{\mathrm{s}}^{j}$ with:
\begin{equation}
\label{eq:kappatilde}
    \tilde{\kappa}_{\mathrm{s}}^{j}
= \sum_{k\neq j} 4\gamma\, g_k^2\, p_k/(4\gamma^2+\delta_{jk}^2)
\end{equation}
and its frequency is also shifted by $\tilde{\delta}_j = \delta_j + \delta_{s}^j$, with
\begin{equation}
\delta_{s}^j = - \sum_{k\neq j} g_k^2\, p_k\,\delta_{jk}/(4\gamma^2+\delta_{jk}^2).
\end{equation}
The latter is antisymmetric in $\delta_{jk}$ and averages to zero for a distribution $\rho_\delta$ symmetric about $\delta_j$; we neglect it in what follows. For a spin at line centre and an ensemble with uniform polarisation ($p_k = \bar{p}$, $2\gamma\ll\Gamma$), integrating over $\rho_g$ and $\rho_\delta$ gives $\tilde{\kappa}_{\mathrm{s}}^{j}\to\bar{p}\,\kappa_{\mathrm{s}}$ with $\kappa_{\mathrm{s}} = 4 g_{\mathrm{ens}}^2/\Gamma$~\cite{staudt2012}, the loss rate that the spins add to the resonator and that we measure in
Fig.~\ref{fig:fig2}a. The spins thus broaden the resonator by an amount proportional to their own polarisation, which slows the Purcell decay down to $\tilde{\Gamma}_{\mathrm{P}}^{j}
= g_j^2\tilde{\kappa}_j/(\tilde{\kappa}_j^2/4 + \tilde{\delta}_j^2)$.

Turning to the third term of Eq.~\eqref{asigma_ff} and re-evaluating the evolution of $\sigma_z^j$ in its presence, keeping the absorptive part of each denominator, we obtain
\begin{equation}
  \label{eq:sigmaz_ff}
  \begin{split}
    &\frac{d}{dt}\langle\sigma_z^j\rangle
    = -\tilde{\Gamma}_{\mathrm{P}}^{j}
       \left[1 + (1+2\bar{n})\langle\sigma_z^j\rangle\right] \\
    & + \frac{\langle\sigma_z^j\rangle}{2}\,
       \tilde{\Gamma}_{\mathrm{P}}^{j}
       \sum_{k\neq j} \frac{2\gamma}{4\gamma^2 + \delta_{jk}^2}
       \left[-2 i g_k\left(\langle a\sigma_+^k\rangle
         - \langle a^\dagger\sigma_-^k\rangle\right)\right] .
  \end{split}
\end{equation}

By Eq.~\eqref{sigma_z_brut_de_pomme} the bracket is nothing but the photon
emission rate $\frac{d}{dt}\langle\sigma_z^k\rangle$ of spin $k$, so that
\begin{equation}
  \label{PurcellBD}
  \begin{split}
 & \frac{d}{dt}\langle\sigma_z^j\rangle
  = -\tilde{\Gamma}_{\mathrm{P}}^{j}
    \left[1 + (1+2\bar{n})\langle\sigma_z^j\rangle\right] \\
  &+ \frac{\langle\sigma_z^j\rangle}{2}\,\tilde{\Gamma}_{\mathrm{P}}^{j}
    \sum_{k\neq j} \frac{2\gamma}{4\gamma^2+\delta_{jk}^2}\,
    \frac{d}{dt}\langle\sigma_z^k\rangle .
    \end{split}
\end{equation}
The kernel $2\gamma/(4\gamma^2+\delta_{jk}^2)$ restricts the sum to the spins lying within the homogeneous linewidth of spin $j$; its integral over $\delta_{jk}$ is $\pi$ irrespective of $\gamma$, so the magnitude of this term is set by the spectral density of spins $N\rho_\delta(\delta_j)$ and not by $\gamma$ itself. The physical content is transparent: a spin $j$ sitting in the ground state of the transition can reabsorb a photon emitted through the resonator by a near-resonant spin $k$ ($\frac{d}{dt}\langle\sigma_z^k\rangle < 0$), which feeds $\langle\sigma_z^j\rangle$ back up. The correction is proportional to $\langle\sigma_z^j\rangle$ and therefore vanishes for an unpolarised transition and grows as population accumulates in $\ket{6}$, which is the regime our DNP protocol creates. Averaged over the ensemble, both this term and the broadening of Eq.~\eqref{eq:kappatilde} act in the same direction: the radiative relaxation is no longer a fixed per-spin rate but a polarisation-dependent, reduced one.

\subsection{Numerical simulations}\label{m:DNPnumerics}
In this section, we present the numerical fit of the DNP data of Fig.~\ref{fig:fig5} using the above description of the Purcell effect. We start by dividing the spin ensemble into $(g,\delta)$ families as explained in Appendix~\ref{m:device}, with a number of spins set according to the spin frequency distribution and coupling distribution corresponding to a spacing of $d = \qty{1.1}{\micro\meter}$. For each sub-family, we consider the 20 possible states, with the 9 ground states being initially equally occupied with identical population.

We then simulate the entire DNP protocol:
\begin{itemize}
    \item We apply the WURST pulses by transferring the populations within a family taking into account the efficiency we estimate for each family in Appendix~\ref{m:WURST} for nuclear and electronic pulses.
    \item We solve for the dynamics of the Purcell relaxation for a duration $t$ taking into account 1)  the intrinsic relaxation effects in the ensemble (see Appendix~\ref{m:intrinsicT1}) considered as family-independent; 2) the Purcell effect as described in Eq.~\ref{PurcellBD} taking into account the renormalization created by the resonator modulation during relaxation, as well as the resonator loading created by the spins, and the spin reexcitation created by inter-family flip-flops.
    \item We stop at time $t$ to extract the spin distribution on the $|6\rangle\leftrightarrow|15\rangle$ transition and compute the expected signal according to our modelling of the CPMG sequence (see Appendix~\ref{m:CPMG}).
    \item To simulate successive iterations of the DNP cycle, we let cycle $i$ evolve to $t=\qty{320}{\second}$ and feed the corresponding spin ensemble state as initial point for cycle $i+1$.
\end{itemize}

We perform this fit with the following parameters as fixed: intrinsic relaxation mechanisms, WURST transfer population matrix, inter-chip distance, and atom-spin efficiency $\eta_I=60\%$, and we introduce only a single overall scaling factor to adjust the numerical CPMG signal to the measured averaged echo amplitude. This simulation is shown in Fig.~\ref{fig:fig5} of the main text.

\begin{figure}[htbp]
    \centering
    \includegraphics[width=\columnwidth]{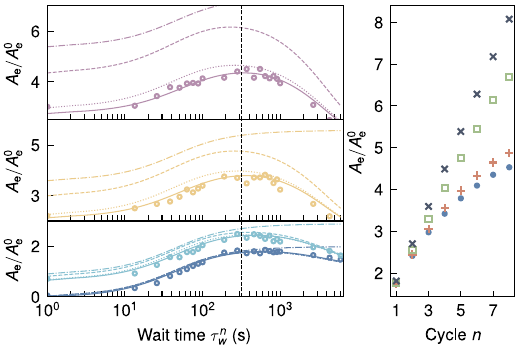}
    \caption{\label{figm:DNPlims} Identifying the limits of our experimental realization of the DNP protocol. We compare four simulation conditions of decreasing completeness: (i) all imperfections included, (ii) no Purcell breakdown, (iii) no Purcell breakdown and no WURST imperfection, (iv) no Purcell breakdown, no WURST imperfection, and no electronic or nuclear intrinsic decay. (Left) Simulated echo intensity after $n$ iterations of the DNP protocol as a function of the waiting time in the $n$th transfer, with all other waiting times fixed at \qty{320}{\second}, for $n = 1, 2, 5, 7$ (blue, cyan, yellow, pink); line style indicates the simulation condition (solid, dotted, dashed, dash-dotted, respectively, for (i)-(iv)). Markers show the experimental data from Fig.~\ref{fig:fig5}d. (Right) Simulated DNP echo enhancement at fixed waiting time $t_W=\qty{320}{\second}$ as a function of cycle number $n$, for the same four conditions, indicated by marker shape ($\circ$, $+$, $\square$, $\times$, respectively, for (i)-(iv)).}
\end{figure}

Using this simulation, we infer the limitations of our DNP scheme, see Fig.~\ref{figm:DNPlims}. Repeating the simulation by cancelling the terms relating to the breakdown of the Purcell effect, we find that our DNP scheme should yield a pumping efficiency of $4.9$, yet their effects on the slow-down of the recovery is increasingly visible when we model the polarisation enhancement as a function of recovery time at higher $n$. The discrepancy to the ideal factor $9$ arises from two other effects : assuming perfect WURST pulses lead to a pumping efficiency of $6.7$. Assuming in addition the absence of hyperfine and electronic decays lead to a pumping efficiency of $8.1$, with the remaining discrepancy due to the fixed pumping time (\qty{320}{s}) compared to the slow decay of spins lying far from the resonator.

\begin{figure}[htbp]
    \centering
    \includegraphics[width=\columnwidth]{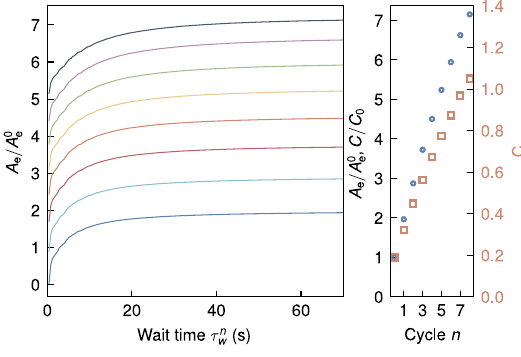}
    \caption{\label{figm:DNPoverachiever} Simulation of the DNP protocol for a 20-$\Omega$ resonator, taking $d=\qty{300}{
\nano\meter}$, and a concentration $20\times$ less.  (Left) Simulated echo intensity after $n = 1, ..., 8$ iterations of the DNP protocol as a function of the waiting time in the $n$th transfer, with all other waiting times fixed at \qty{80}{\second}. (Right) Simulated DNP echo (blue circles) and cooperativity (orange squares) enhancement at fixed waiting time $t_W=\qty{80}{\second}$ as a function of cycle number $n$.}
\end{figure}

To improve the efficiency, it is thus clear that the Purcell effect should dominate more largely over the other intrinsic relaxation processes. We thus simulate for a smaller flip-chip spacing ($d=\qty{300}{\nano\meter}$) and smaller resonator impedance ($20~\Omega$), with a spin ensemble 20 times more dilute, see Fig.~\ref{figm:DNPoverachiever}. We find that in this case the Purcell effect is more efficient (enhancement factor of $7.2$ including Purcell breakdown and hyperfine diffusion), and that the pumping efficiency increases the cooperativity from its initial values of $0.19$ to $1.05$ in 10 minutes, showing that the Purcell breakdown does not prevent reaching unit cooperativity in a reasonable time, for ensembles even more dilute than the one we operate with.

\bibliographystyle{apsrev4-2}
\bibliography{biblio}

\end{document}